# Advances in applications of time-domain Brillouin scattering for nanoscale imaging

Vitalyi E. Gusev[1,a)] and Pascal Ruello[2]

[1]Laboratoire d'Acoustique de l'Université du Maine, UMR CNRS 6613, Le Mans Université, 72085 Le Mans, France

[a)] Author to whom correspondence should be addressed. Electronic mail: vitali.goussev@univ-lemans.fr

[2]Institut des Molécules et Matériaux du Mans, UMR CNRS 6283, Le Mans Université, 72085 Le Mans, France

**Abstract**

Time-domain Brillouin scattering is an all-optical experimental technique based on ultrafast lasers applied for generation and detection of coherent acoustic pulses on time durations of picoseconds and length scales of nanometers. In transparent materials scattering of the probe laser beam by the coherent phonons permits imaging of sample inhomogeneity. The transient optical reflectivity of the sample recorded by the probe beam as the acoustic nanopulse propagates in space contains information on the acoustical, optical, and acousto-optical parameters of the material under study. The experimental method is based on a heterodyning where weak light pulses scattered by the coherent acoustic phonons interfere at the photodetector with probe light pulses of significantly higher amplitude reflected from various interfaces of the sample. The time-domain Brillouin scattering imaging is based on Brillouin scattering and has the potential to provide all the information that researchers in material science, physics, chemistry, biology etc., get with classic frequency-domain Brillouin scattering of light. It can be viewed as a replacement for Brillouin scattering and Brillouin microscopy in all investigations where nanoscale spatial resolution is either required or advantageous. Here we review applications of time-domain Brillouin scattering for imaging of nanoporous films, ion-implanted semiconductors and dielectrics, texture in polycrystalline materials and inside vegetable and animal cells, and for monitoring the transformation of nanosound caused by nonlinearity and focusing. We also discuss the perspectives and the challenges for the future.

## I. INTRODUCTION

Imaging materials and probing their properties (morphology, electrical and magnetic properties for example) at the nanoscale with near-field methods such as Scanning Force Microscopy[1], including all its extensions (Atomic Force Microscopy AFM[2,3], magnetic Force Microscopy MFM[4], piezoelectric Force Microscopy PFM[5], STM-Scanning Tunneling Microscopy[6], Near-field Scanning Optical Microscope NSOM[7] and AFM-based nanotribology tools[8]) has been significantly improved since the nineteen eighties and represents now an array of powerful diagnostics in materials science, in nanotechnology and in biology as well[9]. When the properties or characteristics of interest are not limited to the surface or/and not accessible by the near field methods, i.e., it means when the region to be probed is beneath the surface typically (10 nm-1 $\mu$m), different strategies are required to get the information, most of them being based on the use of a penetrating beam or wave to probe the materials. Electron beams can be used to realize subsurface image of materials by cross-section analysis with atomic resolution through the well-known Transmission Electron Microscopy TEM or with high resolution transmission electron microscopy HRTEM but these techniques require however individualized sample preparation and materials that do not undergo damage under electron beams irradiation[10]. X-Ray Reflectivity (XRR) is also widely used for the studies of thin and ultra-thin films, and with specific algorithms, permits reconstruction of the structure profile of these films[11]. Optical super resolution methods have been recently developed for biological materials imaging including photo-activated localization microscopy (PALM) and stochastic optical reconstruction microscopy (STORM)[12,13]. In general, optical confocal spectrometry such as used in Brillouin and Raman spectroscopies permit subsurface investigation through depth dependent lattice dynamics. But even with latest confocal optical geometry, the Brillouin and Raman spectroscopy based imaging of materials remains limited to micrometer scale[14,15]. Besides electromagnetic waves and electrons or particles beams, the use of penetrating acoustic waves is also very widely used for imaging in both materials science and medicine[16-20] (Fig. 1). The so-called "echography" is based on the analysis of reflection and transmission of ultrasound with various extensions (which is out of the scope of this review). Echography is also well-known in structural health monitoring for investigation at a macroscopic scale and is based on imaging with traditional bias-controlled piezoelectric transducers that are used to emit and receive the sound waves. Standard medical ultrasound imaging gives an in-depth resolution not better than tens of micrometers, because of MHz limited instrumental bandwidth and of the viscous damping of sound at the frequencies higher than a few MHz, thus precluding nanoscale imaging. Nanoscale imaging can be achieved using particular methods of picosecond ultrasonics.

Picosecond ultrasonics, pioneered in the 1980ies[21-25], is a technique that uses ultra-short pump laser pulses to generate coherent acoustic pulses (CAPs) through the optoacoustic effect and uses time-delayed, ultra-short probe laser pulses to detect the CAPs (through the acousto-optic effect) as the CAPs propagate through a transparent medium or reach surfaces or interfaces of an opaque medium. See Fig. 1 (a). A thin film pump-light-absorbing optoacoustic generator for launching the CAPs into transparent samples can be located on the interface of a film and a substrate[26-28] as in Fig. 1 (a). The thin film generator (semitransparent for probe laser pulses) can be also deposited on the front surface of a transparent film or bulk material[23-25,29-31] as in Fig. 5(a). Even a single atom layer light absorber[32] can be used to generate the CAPs. There is no need for additional optoacoustic generators in the case of a pump light absorbing substrate, film or bulk material[26,33-35] as in Fig. 6(a). The mechanism of the optoacoustic transformation is mainly thermo-elastic, i.e., deformation of the material by stresses induced by laser-induced heating[22,36]. The other mechanisms, such as electron-hole-phonon deformation, the inverse piezoelectric effect, electrostriction, etc., can be efficient under particular circumstances[22,36-38]. The duration of the CAP depends, in general, on the duration of the pump laser pulses, the life time of the photo-induced mechanical stresses and the time of sound propagation across the localization depth of the photo-induced mechanical stresses, which, in turn, depends on the pump light absorption depth[22,36,37].

In the case where sound generation takes place in an opaque (metallic) thin film attached to the sample, as shown in Fig. 1 (a) and 5(a) the thickness and absorption of the thin film determine the acoustic pulse length. Through optimization of the optical and material factors involved in producing the acoustic wave, it is possible to generate coherent acoustic pulses of picoseconds duration with spatial lengths of nanometers making them suitable for nanoscale imaging.

In application to transparent materials, the technique of picosecond ultrasonics was originally referred to as picosecond acoustic interferometry (PAI)[23-25]. In optically transparent materials (see Fig. 1) the probe laser pulse preferentially interacts with those GHz-frequency phonons in the CAP spectrum that satisfy the momentum conservation law for photon-phonon, photo-elastic interaction, that is, they satisfy the Brillouin scattering (BS) condition[22,39,40]. The experimental method is based on heterodyning where weak light pulses scattered by the coherent acoustic phonons interfere at the photodetector with probe light pulses of significantly higher amplitude reflected from various interfaces of the sample (Fig. 2(a)).

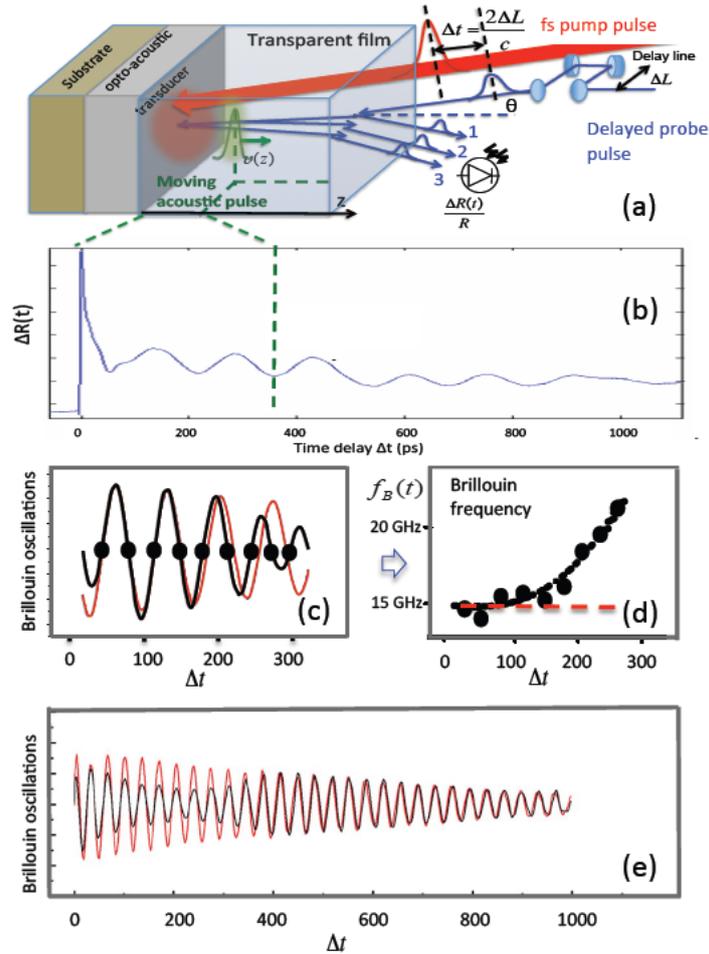

**FIG. 1.** Principles of opto-acousto-optic depth profiling by picosecond acoustic interferometry. (a): A thin film pump-light-absorbing optoacoustic generator for launching the coherent acoustic pulses into transparent samples is located on the interface of a film and a substrate. The transient optical reflectivity signal (b) contains a sinusoidal oscillating component, the Brillouin oscillation. The Brillouin frequency in an inhomogeneous film can change with time[28]. In the left part of (c) a comparison of the experimental Brillouin oscillation, in black, with model constant-frequency Brillouin oscillation, in red, reveals an increase in the experimental Brillouin frequency, also evidenced by the Fourier transform in a moving time window in the right part of the figure. Reproduced with permission from ACS Nano **6**, 1410 (2012). Copyright 2012 American Chemical Society. In (d) the Brillouin oscillation amplitude in an inhomogeneous medium, in black, exhibits variations in addition to those caused by the probe photons and the CAPs absorption in homogeneous media, shown in red[42], while the Brillouin frequency is fairly constant. Reproduced from Appl. Phys. Lett. 94, 111910 (2009), with the permission of AIP Publishing.

The recorded transient optical reflectivity signal is proportional (to first order) to the product of the electric fields of the two light fields. Thus, heterodyning of a weak field against a strong one is achieved. The measured signal varies with time because the relative phase of the light scattered by the CAP and reflected by immobile interfaces continuously changes with time due to the variation in the spatial position of the CAP. If the CAP propagates at a constant velocity in a spatially homogeneous medium, the phase difference between the interfering light fields changes in time linearly causing sinusoidal variations in the signal amplitude at a frequency precisely equal to the Brillouin frequency (BF)[22-25] (See Fig. 1 (b) and Section II). The acoustically-induced oscillating contribution to the reflectivity signal (see Fig. 1 (c) and (d)) is commonly known as the Brillouin oscillation (BO). In view of the above mentioned intrinsic relations of the PAI to the Brillouin scattering the technique is currently more and more often called the time-domain Brillouin scattering (TDBS). In the collinear scattering geometry, the BF depends linearly on the product of the sound velocity and the optical refractive index of the sample at the probe wavelength. The potential of the PAI for depth-profiling of the materials has been explicitly pointed out in Ref. 24, viz., "Velocity variations as a function of distance from the surface can be determined from a measurement of how the period of oscillation varies with time delay of the probe pulse". Surprisingly though, the first experimental application of the PAI for depth-profiling was published almost 20 years later[41]. At the same time, it was shown experimentally[42] that the information on material spatial inhomogeneity can be revealed by monitoring the evolution in time not of the frequency of the BO, but rather its amplitude. Indeed, the amplitude of the acoustically-induced transient reflectivity signal is proportional to the strength of the acousto-optic interaction, that is, to the photo-elastic constant characterizing variation of the optical properties of the probed media with acoustic strain[39,40]. An inhomogeneity in the material photo-elasticity could be the dominant mechanism providing the contrast for depth-profiling by the TDBS[28,42]. Following the first experiments on depth-profiling of ion-implanted crystals[42] and nanoporous thin films,[41] progress was reported in both of these applications[28,43-46] and extended to imaging of radiation damage[43-45]. More recently the TDBS was applied to two and three-dimensional imaging of textured polycrystalline materials at high pressures[47,48] and inside both plant and animal cells[49-51]. Very recently the TDBS was used to reveal the gradient of the temperature distribution in liquid[52]. The imaging was based on analysis of the temporal evolution of either the BF [47,50,52] (Fig. 1 (c)) or the BO amplitude[43-46] (Fig. 1 (d)), or of both[28,48,49]. An axial resolution of about 40 nm was achieved in low-k nanoporous sub-$\mu m$ thick films[28,41] and of about 30 nm in ion-irradiated semiconductors[42,43].

Among other imaging techniques based on the combination of coherent acoustical and optical waves it is worth mentioning for comparison the optoacoustic (photoacoustic) microscopy[53-60] applied for imaging at mm/$\mu m$ spatial scales. In optoacoustic imaging coherent acoustic waves, thermo-elastically generated by the optically-induced heating of spatially inhomogeneous media, read the information on the inhomogeneity and deliver it to the detection modality. The contrast is primarily due to spatially inhomogeneous optical absorption. Photoacoustic microscopy (PAM) and optical-resolution photoacoustic microscopy (OR-PAM) have seen their widest application in biology and medicine[55-60]. The physical principles of TDBS imaging and the origin of the contrast in TDBS imaging are drastically different from those in optoacoustic (photoacoustic) imaging. In TDBS imaging coherent ultrashort acoustic pulses travel in the inhomogeneous medium, while probe light pulses read information on the material parameters at the current position of the coherent acoustic pulse and deliver it to the detection modality. The contrast in PAI imaging is usually a result of the spatial inhomogeneity of optical properties (refractive index), acoustical properties (sound velocity and acoustic impedance) and/or opto-acoustic properties (photoelastic tensor).

Speaking generally, TDBS imaging is based on Brillouin scattering and has the potential to provide all the information that researchers in materials science, physics, chemistry, biology etc., get with classic frequency-domain Brillouin scattering (FDBS) of light. However, TDBS incorporates Brillouin scattering not by the thermal phonons that exist at all points of the sample, but rather by the coherent phonons of much higher amplitudes intentionally launched into and highly localized in the sample. Thus, in addition to its reliability and speed for three dimensional imaging of transient processes at nanoscale, TDBS can be viewed as a replacement for Brillouin scattering and Brillouin microscopy[14,15] in all investigations where nanoscale spatial resolution in materials characterization is either required or advantageous (Fig. 2).

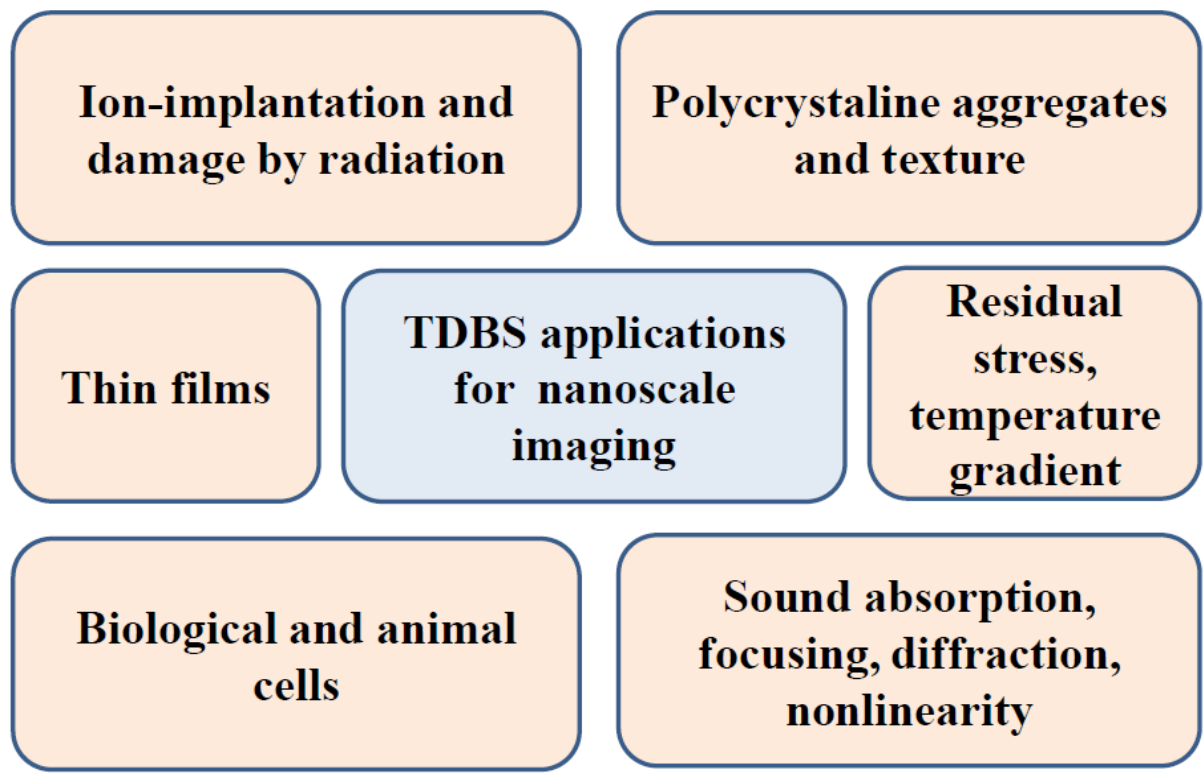

**FIG. 2.** Applications of time-domain Brillouin scattering for imaging.

Below we review some research results constituting the background for the progress achieved in TDBS imaging and the results of several imaging experiments. We also discuss the applications of the TDBS for monitoring of nonlinear acoustic processes[61,62] and hypersound nanofocusing[63] as well as perspectives and the challenges for the future.

## II. THEORETICAL BACKGROUNDS FOR IMAGING VIA TIME-DOMAIN BRILLOUIN SCATTERING

In time-domain Brillouin scattering (TDBS) probe laser pulses are scattered by laser-generated picosecond coherent acoustic pulses (CAPs) due to the acousto-optic effect[22,39,40]. By monitoring the reflected light, information on the time development of the CAPs is obtained as they travel through the experimental sample[22]. In the most of the experiments the transient relative changes $dR(t)/R$ in the intensity of the reflected probe light are detected. The reflectivity $R$, and electromagnetic field reflectivity are related by $dR(t)/R = 2\operatorname{Re}[dr(t)/r]$. Among possible contributions to this signal there is one generated by the acousto-optic interaction, which is proportional to the photo-elastic (acousto-optic) coefficient and the overlap integral of the acoustic strain field with the optical field sensitivity function[22]. The integral is over the complete volume probed by the light, while the sensitivity function depends on the specific geometry of the light distribution in the sample in the absence of the CAPs[22,64,65]. The integral gives the solution for electromagnetic wave propagation equation in an optically inhomogeneous medium under the single-scattering approximation[66-68], which is valid because the perturbation of optical properties by the CAP is weak. In particular, when light is normally incident on the sample surface probes the propagation of the acoustic pulse in an optically homogeneous half space (Fig. 3), the photo-elastic contribution is given by

$$\frac{dr}{r} = -iC \int_0^\infty \eta_{zz}(t,z) e^{i2kz} dz , \qquad (1)$$

where $C$ is the known complex constant proportional to the acousto-optic interaction[22,64,65], $\eta_{zz}(t,z)$ is the strain field of the longitudinal CAPs propagating along the $z$ axis, and $k$ is the complex wave number of the probe light in the medium. Thus, $dr(t)/r$ provides information on the evolution in time of the Laplace component of the acoustic field with the complex-valued Laplace parameter $p = -i2k$ [64].

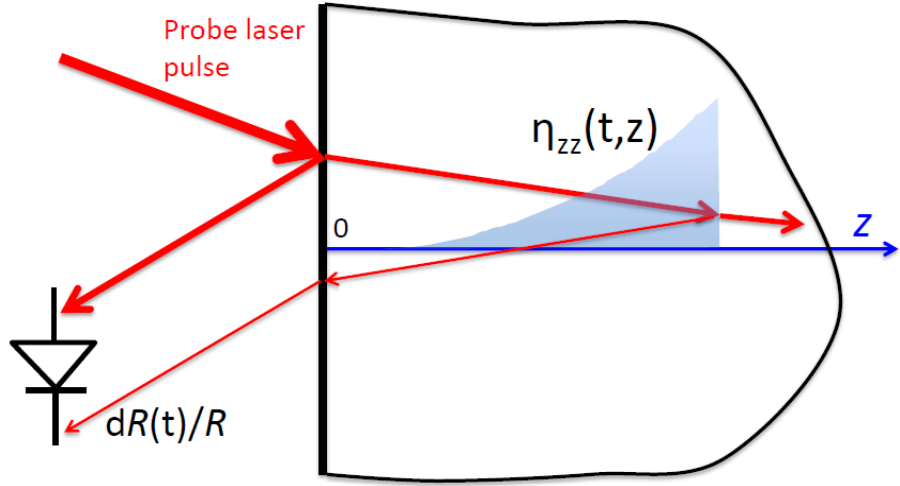

**FIG. 3.** Time-domain Brillouin scattering in a half space. The laser beams are shown at an angle for clarity. Reproduced from J. Appl. Phys. 110, 124908 (2011), with the permission of AIP Publishing.

Other interpretations of Eq. (1) can be given, which could more insightful for the physical interpretation of specific experimental realizations. For the TDBS imaging it is essential that the strain field $\eta_{zz}(t,z)$ in Eq. (1) at each time instant is strongly localized in space in form of a nanoscale CAP. For example, for a CAP propagating in the positive z direction starting at t=0 from z=0 (Fig. 3), slow evolution of its strain field in a medium with inhomogeneity of the acoustic parameters only can be written in the form

$$\eta_{zz}(t,z) \equiv \eta_{zz}(t, \xi = z - z_a(t)) = \int_{-\infty}^{+\infty} \tilde{\eta}_{zz}(t,q) e^{-iq\xi} dq/(2\pi). \qquad (2)$$

Here $z_a(t) \equiv \int_0^t v_s(t') dt'$ describes the variation of the CAP position in space, $v_s$ is the velocity of the CAP, which depends on the co-ordinate and, thus, is different in the different time instants of CAP propagation, while the first argument of the strain function describes possible slow evolution of the CAP's amplitude and profile in the time domain or of the components of the CAP spectrum $\tilde{\eta}_{zz}(t,q)$ in the reciprocal q space. Slow evolution of the CAP takes place if the width/length of the CAP spatial localization region, $l_a(t)$, is significantly shorter than a characteristic spatial scale for the variation of the acoustical parameters of the medium and the characteristic lengths of dispersion, diffraction and nonlinearity[36,37,67-69]. However, here for the simplicity we completely neglect possible transformation of the CAP, which could be caused by dispersion, diffraction and nonlinear phenomena, in order to highlight the manifestations of the medium inhomogeneity. A detailed derivation of the solution for $dr/r$ when the medium is also optically inhomogeneous and the propagation of probe laser radiation and a coherent acoustic pulse are treated within the framework of geometrical optics and geometrical acoustics[67,68], respectively, can be found in Ref. 65. Substitution of the strain field (2) into Eq. (1) leads, after the change of the integration variable, to

$$\frac{dr}{r} = -iCe^{i2kz_a(t)}\int_{-z_a(t)}^{+\infty}\eta_{zz}(t,\xi)e^{i2k\xi}d\xi \cong -iCe^{i2kz_a(t)}\int_{-l_a(t)/2}^{l_a(t)/2}\eta_{zz}(t,\xi)e^{i2k\xi}d\xi. \quad (3)$$

The final form of the right-hand-side in Eq. (3) is derived under the assumption that the propagation distance of the CAP exceeds its spatial length, $z_a(t) \geq l_a(t)/2$. The solution in Eq. (3) is very instructive. First, it demonstrates that quasi-periodic oscillations of $dr/r$ in time domain are caused by the time-dependent phase shift, $\varphi_D = 2kz_a(t) = 2k\int_0^t \upsilon_s(t')dt'$, of the probe light scattered by the CAP. This is just a classic Doppler phase shift of the probe light pulses when the ultrafast laser is operating as radar to monitor the motion of the acoustically-induced weakly reflecting mirror. The derivative of this phase shift over time provides the Doppler frequency shift of the probe light, $\omega_D = \partial\varphi_D/\partial t = 2k\upsilon_s(t)$. It is worth noting that neither $\varphi_D$ nor $\omega_D$ depend on the acoustic pulse profile (spatial spectrum). Under the assumed conditions all the points of the CAP strain profile are propagating with the same local velocity, corresponding to the current spatial position of the CAP, while $\varphi_D$ and $\omega_D$ themselves do not contain any indication on the preferential scattering of the probe light by a particular spectral component of the CAP. These theoretical predictions are independent of the $l_a(t)$ magnitude relative to the characteristic scale $\pi/k \equiv \lambda/2$ of the variations of the sensitivity function, $f(x) \propto \exp(i2kx)$, in Eq. (1). However, in the TDBS imaging experiments the CAPs are dedicatedly generated of the nanoscale length $l_a(t)$ significantly shorter than the wavelength of the probe optical light in order to achieve imaging with sub-optical wavelength resolution. In this case, assuming $l_a(t) << \lambda/2$, Eq. (3) can be approximated by

$$\frac{dr}{r} \cong -iCe^{i2kz_a(t)} \int_{-l_a(t)/2}^{l_a(t)/2} \eta_{zz}(t,\xi)(1+2ik\xi+...)d\xi \tag{4}$$

The solution in Eq. (4) demonstrates that in leading order the signal is proportional to the strain integrated over the localization region of the acoustic pulse (total "area" $S(t)$ of the strain pulse), i.e., to the displacement of the medium retained in it after the CAP propagation. Second term in Eq. (4) should be taken into account in the case of the symmetric bipolar CAPs of zero surface area, which could be generated by the pump laser pulses near the mechanically free surface of the medium[22,36,37]. Equation (4) demonstrates that in homogeneous medium ($z_a(t) = \upsilon_s t$, $S(t) = S_0 = const$) the TDBS signal, $dr/r \cong -iCS_0 \exp(i2k\upsilon_s t) \propto f(x=\upsilon_s t)$, can be interpreted as an imaging of the spatial sensitivity function, while weak inhomogeneities of the medium could be revealed because they are causing deviations from this "background" image.

The fact that the probe light exhibits maximum scattering efficiency by the acoustic phonons with the wavenumber $q = 2k \equiv 2k_p$ is evidenced, when the spectral presentation of the CAP from Eq. (2) is substituted in Eq. (3) and the integration over the spatial coordinate is done:

$$\frac{dr}{r} \cong -iCe^{i2kz_a(t)} \frac{1}{\pi} \int_{-\infty}^{\infty} \tilde{\eta}_{zz}(t,q) \frac{\sin[(q-2k)l_a(t)/2]}{(q-2k)} dq. \tag{5}$$

For the qualitative interpretation of Eq. (5) a precise definition of the characteristic dimension $l_a(t)$ of the CAP localization is not necessary. The earlier assumed inequality, $l_a(t) << \lambda/2 = \pi/k$, is sufficient to estimate that a broad spectrum of the acoustic wavenumbers contributes to scattering of the probe light in Eq. (5). The characteristic width of this spectrum, defined by $|q - 2k| \leq 2\pi/l_a(t) >> 2k$, significantly exceeds $2k$. Thus, although the TDBS scattering exhibits the maximum sensitivity to the spectral component $\tilde{\eta}_{zz}(t, 2k)$ it is not a narrow band process and most of the CAP spectral components are efficient in light scattering to some extent. This broadening of the spectrum of coherent phonons that could contribute to light scattering with comparable efficiency is due to the smallness of interaction length of probe light with the coherent phonons that are carried by a nanoscale CAP. The relation $q = 2k$ is a well-known result of the momentum conservation in back-scattering of a photon by a phonon[40,66]. However, the small interaction length does not provide opportunity for the synchronous interaction of probe light with the $q = 2k$ phonons to "win" significantly relative to light scattering by other phonons.

Because of the relation $q \equiv \omega/\upsilon_s$, where $\omega$ is the cyclic frequency of the phonon, it is common to say that TDBS predominantly monitors the slow time dynamics of the spectral Fourier component of the CAP at the frequency $\omega = 2k\upsilon_s \equiv \omega_B$, known as the Brillouin frequency (BF) or the BF shift of a photon back-scattered by a phonon. The BF shift coincides with the Doppler frequency shift discussed earlier. Because of this, the oscillating TDBS signals, due to the time-dependent phase shift of the scattered light, are commonly called the Brillouin oscillations (BO). The wave number $q = 2k \equiv 2k_p$ and the wavelength $\lambda = 2\pi/q$ are known as the Brillouin wave number, $q_B$, and the Brillouin wavelength, $\lambda_B$, respectively.

In Ref. 69 it is demonstrated that the dominant contribution to the TDBS signal comes from the sharp fronts in the strain profile of the CAPs, as it is symbolically depicted in Fig. 3. This result is just a different formulation of the fact that light is not scattered inside a homogeneously strained medium. Thus, the spatial resolution of TDBS imaging can be better than the total spatial length of a CAP.

The separation in Eq. (2) of the amplitudes and the phases of $C$ and $\tilde{\eta}_{zz}(t,\omega)$, i.e., $C = A_C \exp(i\varphi_C)$, $\tilde{\eta}_{zz}(t,\omega_B) = A_\eta(t)\exp[i\varphi_\eta(t)]$, leads to[69] $dR(t)/R = 2A_C A_\eta \sin(\omega_B t + \varphi_\eta + \varphi_C)$. Thus the $dR(t)/R$ signal oscillates at a frequency given by

$$\omega_{dR/R} = \partial(\omega_B t + \varphi_\eta + \varphi_C)/\partial t = \omega_B + \partial\varphi_\eta/\partial t, \qquad (3)$$

which, in general, differs from the BF defined above, i.e., $\omega_{dR/R} \neq \omega_B$. For example, when TDBS is applied for the monitoring of the nonlinear CAP propagation, $\omega_{dR/R} \neq \omega_B$ provides information on the velocity of the weak shock front in the profile of the CAP[61,69].

In conclusion of this theoretical Section we would like to attract attention of the readers to the fact that the TDBS imaging is neither optical nor acoustical imaging, but is acousto-optical imaging in a sense, that the signal is due to the existence of an overlap region of the optical probe and acoustical fields, Eq. (1), and TDBS imaging is an imaging of their interaction. This provides an explanation for why the TDBS imaging can achieve the spatial resolution much higher than that limited by the wavelength of the optical probe light or by the wavelength, $\lambda_B \equiv 2\pi\upsilon_s/\omega_B$, of the acoustic spectral component at Brillouin frequency, $\omega_B$. The above presented theoretical formulas indicate, that, although the probe light is interacting the most efficiently with a particular component of the CAP spectrum, with the wavelength equal to $\lambda_B$, this interaction takes place in the limited volume, which in optically transparent media is limited by the length of the CAP, $l_a$, and can be localized even in narrower region(s) corresponding to the sharp front(s) of the CAP. We denote here the length of the CAP front by $l_f$. Even $l_a$, not speaking on $l_f$, can be much narrower than $\lambda_B$. Two regions in the medium separated by a distance $l$, which is larger than $l_a$ but much shorter than $\lambda_B$, can be theoretically resolved by the TDBS imaging because they are probed not simultaneously, but separately, i.e., one by one, when the CAP is traversing the medium. This ability of the TDBS imaging is due to the application of ultra-short laser pulses. The ultra-short pump laser pulses provide opportunity to generate the CAPs on length scales of nanometers, while the time-delayed ultra-short probe laser pulses interact with the propagating CAPs quasi-instantaneously inside the volume currently occupied by the CAP. As a consequence, there is no interference in the TDBS imaging signal of the contributions coming from the regions separated by $l \geq l_a$. The time-dependent frequency of the Brillouin oscillation in continuously inhomogeneous medium, $\omega_B(t) \equiv \omega_D(t)$, although it contributes to the argument of the oscillating function in Eqs. (2) and (3), is a local parameter of the material at the scale of $l_a$ or shorter scale of $l_f$, as far as $l_a \leq \lambda_B(t)$. As demonstrated by Eqs. (3) and (4) this parameter is determined independently of the spectrum of the acoustic pulse and is, thus, independent of the scale imposed by $\lambda_B$. The acoustic pulse of any profile will cause the same $\omega_B(t) \equiv \omega_D(t)$ when all the points of the profile are propagating at the same velocity. The complementary theoretical argument in favour of the TDBS spatial resolution superior to one limited by $\lambda_B$ is the sensitivity of the TDBS to a large spectrum of the acoustic waves, including those at frequencies much higher than $\omega_B(t) \equiv \omega_D(t)$, as it is demonstrated by Eq. (5). As a consequence, the interpretation of TDBS as the scattering by the localized at the nanoscale moving strain gradients is closer to the physical reality than the interpretations based on the dominant or even predominant role of the BF phonons. Equation (5) explicitly describes a trade-off between the spatial and wave number resolution in TDBS.

The ultimate/highest spatial resolution in determining the parameter $\omega_B(t)$ is theoretically limited by $l_a$ or $l_f$ and theoretically does not require even a registration of at least a single period or a half of the period of the Brillouin oscillation. How to achieve this highest spatial resolution, if it is required in the experiments, by an appropriate processing of the TDBS signals is another question. Qualitatively speaking, as the evaluation/fitting of the signal is commonly achieved in some moving time window of the duration $\tau_w$, then the spatial resolution in the imaging of the BF parameter, $\omega_B(t)$, could be limited by the length $\upsilon_s \tau_w$ when $\upsilon_s \tau_w \geq l_a$ or $l_a \geq \upsilon_s \tau_w \geq l_f$. Some restrictions on the choice of the shortest $\tau_w$ and some other limitations of the TDBS imaging technique will be discussed later in Section V.

## III. APPLICATIONS OF TIME-DOMAIN BRILLOUIN SCATTERING FOR NANOSCALE IMAGING OF INHOMOGENEOUS MEDIA

### A. Monitoring the transmission of coherent nano-acoustic pulses across spatially localized inhomogeneity

The first experiments indicating potential of the TDBS for depth profiling of continuously spatially inhomogeneous materials were those where the transmission of acoustic pulses through the interface between two different optically transparent homogeneous media was monitored[29-31,70-73]. Insightful experiments were performed with SiO$_2$ films deposited on Si substrates[29-31,70,74-77] as presented in the schematic given in Fig. 4.

The experiments[30] demonstrated remarkably large sensitivity of the transient optical reflectivity signals to the optical wavelength of femtosecond laser pulses applied to probe the CAPs in the samples. Figure 4 shows that, depending on the probe wavelength, transient reflectivity signals $dR(t)/R$ can reveal either only the return of the weak acoustic pulse reflected at the SiO$_2$/Si interface to the front surface of the sample

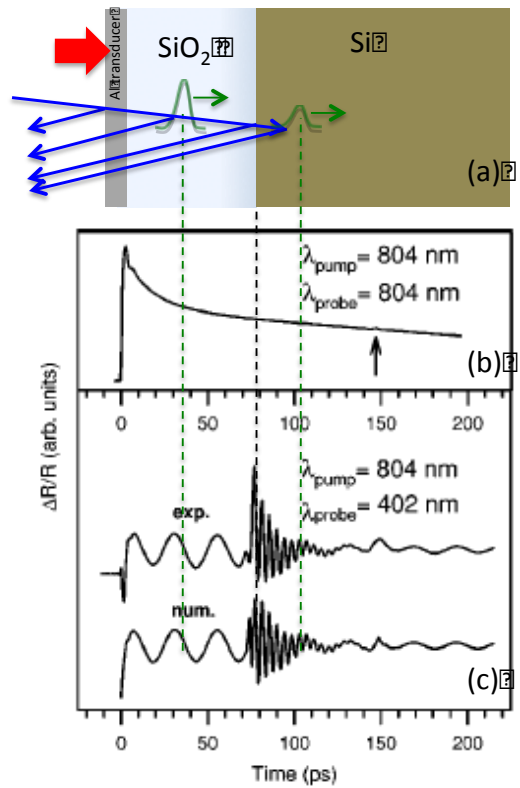

**FIG. 4.** The detection of a spatially localized inhomogeneity. (a) A metallic, semi-transparent transducer was used to generate the coherent acoustic pulses (adapted from Ref. 30). (b) Reflectivity measurement on an Al/SiO$_2$/Si sample with pump and probe both centered at 804 nm. (c) Reflectivity measurement when the probe is changed for 402 nm and numerical results which well reproduce all the observed oscillations. The Brillouin oscillation amplitude abruptly increases starting near 75 ps when the coherent acoustic pulse reaches the SiO$_2$/Si interface. Reproduced from J. Appl. Phys. 104, 123509 (2008), with the permission of AIP Publishing.

(red probe detection of acoustic echo indicated by an arrow in the middle part of Fig. 4), or a much higher in amplitude BO signal corresponding to monitoring of the CAP first in the SiO$_2$ and then later in the Si (the case of blue probe in the lower part of Fig. 4). It was revealed in Ref. 31 that TDBS signals strongly depend on the probe optical wavelength near the critical points of the silicon energy band structure where there are important variations in the optical parameters of this material but also, and most importantly, in the magnitude of photo-elastic constant. Later the TDBS experiments were realized with other combinations of transparent materials such as various transparent layers deposited on silicon[31,70,74,78], GaSb-GaAs heterostructures[72], ZnO and $SiO_2$ deposited on GaAs,[35] the vacuole of a cell deposited on a thin polyethyleneimine film[79] and even microcapsules composed of polymer shell encapsulating a liquid core[80,81]. The TDBS with a variable probe wavelength were conducted in GaAs[72,73]. In Ref. 82 the TDBS was applied to follow CAP propagation across a grain boundary. One of the most insightful reports[73] indicating the possibilities of TDBS for depth profiling showed how the doping profile in different *n*-doped GaAs homoepitaxial layered structures (see upper part of Fig. 5) could be measured. The authors used an optical pump-probe technique called asynchronous optical sampling (ASOPS) introduced earlier by them in picosecond acoustic studies[83,84]. This technique permitted detection of changes in optical reflectivity over a 1 ns time delay with a signal-to-noise ratio of $10^{-7}$ and 100 fs time resolution in approximately 1 min of acquisition time. It was shown that the doping profile with doping densities of the order of $10^{18}$ $cm^{-3}$ can be detected due to differences in the optical and acousto-optical parameters of the doped and non-doped regions. The experimentally recorded BO for the samples with the doped GaAs film on top of the substrate (Fig. 5) clearly show a behavior substantially different from those as shown in Fig. 4. This difference comes from the fact that for the two experiments the coherent acoustic fields generated by the pump laser pulses differ and have dissimilar spatio-temporal patterns. The shortest acoustic transients are generated in the regions of largest gradients of the photo-induced stresses[36,37], which are localized near the mechanically free surface of the sample and at the interface between the layer and substrate[73,85]. Thus, the CAPs containing BFs are launched at the surface in the direction of the interface and at the interface in two opposite directions. The BO in the middle and lower parts of Fig. 5. corresponds to simultaneous monitoring by the TDBS of three CAPs. However the BO monitoring still performs adequately for determination of interface localization and for comparing the parameters of interfacing materials. Of parenthetic note is that the research papers cited in this Section point out the importance of optical probe color choice in the TDBS depth-profiling experiments.

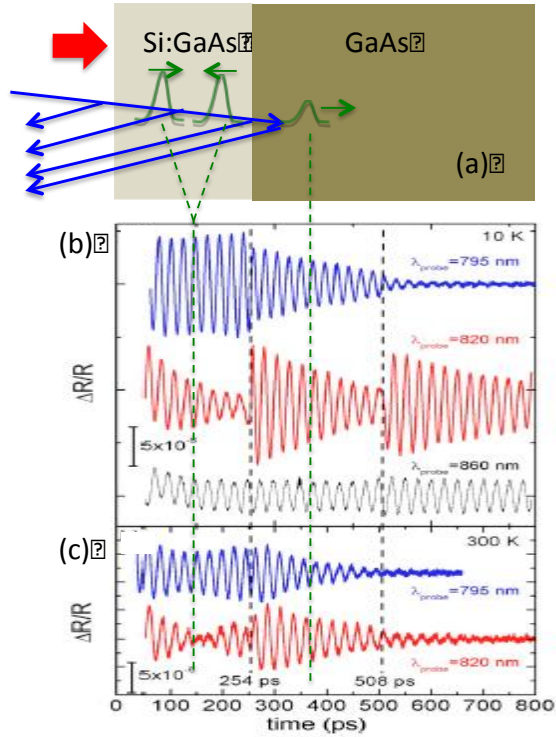

**FIG. 5.** The detection of a spatially localized inhomogeneity. (a) The sample consists of *n*-doped GaAs films on non-doped GaAs substrates (adapted from Ref.73). *dR/R* results were obtained at 10 K (b) and 300 K (c). The measurements were performed with a pump wavelength of 800 nm and probe wavelengths of 795 and 820 nm. *R(t)* transients exhibit a distinct phase shift and with most of them showing an amplitude change at 254 ps, the time a coherent acoustic pulse needs to travel once through the film. Reproduced from Appl. Phys. Lett. 94, 111910 (2009), with the permission of AIP Publishing.

B. Depth-profiling of ion-implanted semiconductors and dielectrics

The first application of TDBS for depth-profiling of the samples with continuously distributed in space inhomogeneity of acousto-optic parameter was reported in Ref. 42. The experiments were carried out using GaSb/GaAs heterostructures (insert in Fig. 6). Defect concentrations in GaAs specimens were created through $He^+$ ion implantation over a wide range of doses, reliably creating damage profiles, which can be simulated using the transport of ions in matter (TRIM) code[86].

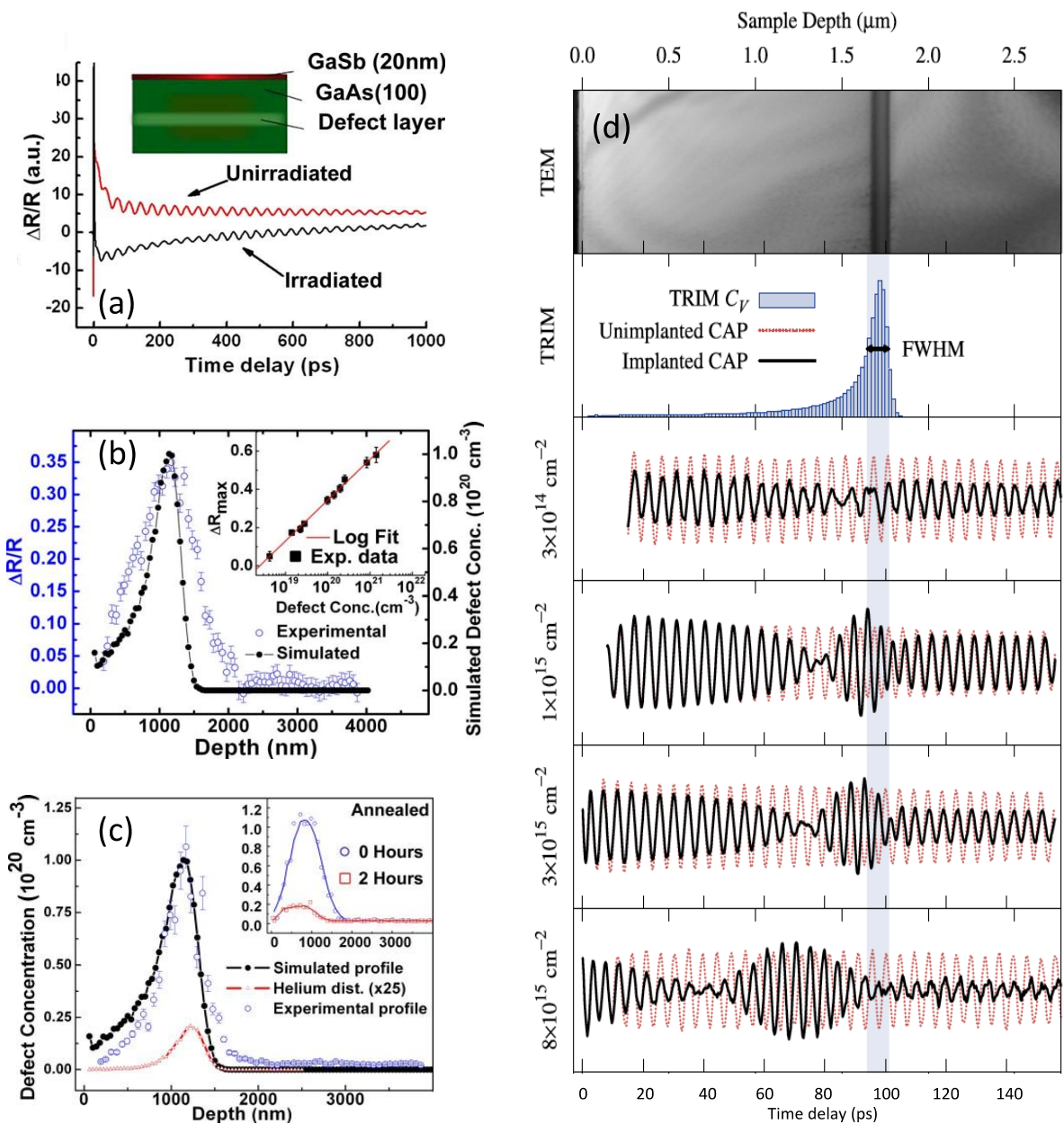

FIG. 6. First experiments on depth-profiling by time-domain Brillouin scattering of a continuously distributed in space acousto-optical inhomogeneity of material. (a) Time-resolved pump-probe optical response of ion-implanted GaAs for undamaged and damaged samples[42]. The coherent acoustic pulses were generated following the interband absorption of pump laser pulses in GaSb layer. The minimum point in space in the amplitude of the Brillouin oscillation in the damaged sample corresponds quantitatively to the position maximum of the defect concentration evaluated by the TRIM numerical code. (b) Difference in oscillatory amplitudes between irradiated sample and undamaged samples, with comparison to the simulated peak defect concentration (solid black curve). Insert: Experimental dependence of peak amplitude modulation as predicted from TRIM calculations which can be used to transform the raw data in (a) into absolute defect concentration profiles. (c) Experimental defect concentration profile for a sample irradiated at 325 keV at a dosage of $7 \times 10^{13}$ ions/cm$^2$. Observed peak defect concentration agrees well with simulated profiles for total damage (black) and helium ion distribution (red, enhanced 25x) from the TRIM code. Insert: Experimentally measured defect profiles before (blue) and after (red) 2 h of thermal annealing at 300 °C. Reproduced from Appl. Phys. Lett. 94, 111910 (2009), with the permission of AIP Publishing. (d) BO in the ion-implanted diamond specimens at multiple fluences (dark lines)[44]. The dotted line behind each curve is the corresponding response for an unimplanted specimen. Above the oscillating responses are the TEM cross-sections for the specimen dosed at $3 \cdot 10^{16}$ cm$^{-2}$ (bright field, two-beam diffraction, g¼(220)) and the damage-induced vacancy distribution as calculated by the TRIM code. The vertical bar indicates the full width at half maximum of the TRIM vacancy profile. The spatial extent of the implantation effects in the diamond specimens on the BO is much broader than seen in the visible structural changes to the lattice shown by TEM. The existence of the node in the amplitude of the BO around which the phase of the oscillations changes by 180 degrees indicates that for threshold values of vacancy concentration the photo-elastic constant of the sample changes its sign passing through its zero value. Reproduced from Appl. Phys. Lett. 94, 111910 (2009), with the permission of AIP Publishing.

The experimental results illustrated in Figs. 1 (d) and 6 (a) demonstrate that the BO in both the undamaged and damaged samples exhibit amplitude attenuation as a result of natural attenuation of the probe light in bulk GaAs, the period and phase remaining identical for both samples, with a distinct difference seen only in the BO amplitudes, which were found to be as much as 35%. The amplitude modulation in the damaged region can be wholly attributed to changes in the photo-elastic constants of GaAs[42]. In order to develop a tool for quantitative characterization of the defect spatial profiles, the authors calibrated the TDBS measurements. It was shown that the raw data, such as those in Fig. 6 (b), can be transformed into quantitative, depth-dependent defect concentration profiles (Fig. 6 (c)). It was also demonstrated that the TDBS may be used to monitor the decrease in defect concentration in the annealing process (Fig. 6 (c)). Note, that the time delay of the pump/probe reflectivity signal can be correlated with crystal depth by simple multiplication of the delay time by the longitudinal speed of sound. This simple procedure enables the features in the reflectivity signal to be mapped directly against the damage-induced vacancy concentration profile as calculated by TRIM (see Fig. 6 (b,c)). As predicted by simulation and verified by ion channeling analysis,[43] the reduction of the TDBS signal amplitude was directly related to the structural damage density.

TDBS with an optical probe frequency near the band-edge is also very sensitive to damage-induced modification of the optical properties of diamond crystals[44]. The bottom four panels of Fig. 6 (d), adapted from Ref. 44, show TDBS signals for diamond specimens irradiated with ion fluencies ranging from $3 \cdot 10^{15}$ to $3 \cdot 10^{16}$ cm$^{-2}$. The regions of strongest ion-induced damage manifest themselves by the BO that are opposite in phase relative to those in the undamaged regions. The front boundary of the zone where the damage is easily distinguishable, close to the node in the amplitude of the BO in Fig. 6 (d), approaches the sample surface with increasing implantation dose. Recently TDBS has been applied for the determination of ion-implantation densities in SrTiO$_3$[45] and 4H-SiC[46].

The totality of experimental results and theoretical interpretations[42-46] convincingly demonstrate that TDBS is a sensitive, noninvasive, and nondestructive tool for studying the depth-dependence of optical properties induced by doping, ion-implantation, or radiation-induced damage. TDBS provides insight into the influence of defects on optical properties and into the specific relationships between the structural modifications of the crystal lattice and the resulting modulation of local optical properties of relevance to the fabrication of various photonic and optoelectronic devices. As reported in Refs. 42-44, the technique "spans at least four orders of magnitude in defect concentration, and has a depth range of the order of tens of micrometers and $\propto$ 10-30 nm depth resolution throughout the probed range". The sensitivity of TDBS makes it a robust defect detection technique with the demonstrated depth limits surpassing those achieved using conventional medium energy ion scattering (MEIS) and ion channeling measurements, which, notably, create damage as a result of analysis.

C. Depth-profiling of nanoporous thin films

The first application of TDBS for the depth-profiling of a continuously distributed elastic inhomogeneity of a sample was reported in Ref. 41. The samples were low dielectric constant, nanoporous organosilicate (SiCOH) thin films (Fig. 7 (a)). Because of their porosity, such materials are mechanically weak. Significant efforts are therefore being directed to improving their mechanical properties by post-deposition treatments. Ultraviolet (UV) radiation-assisted thermal curing, which combines photochemical processing with thermal activation, is perhaps the most promising treatment because it permits efficient removal of porogen, significantly improving mechanical properties with minimum damage to the low-$k$ matrix. However, the intensity distribution of UV light exhibits a gradient inside the low-$k$ film and may create a gradient in its mechanical properties. The first experiment[41] using TDBS for probing a continuously distributed elastic inhomogeneity was conducted using a single-wavelength ($\propto 800$ nm) and an optical configuration with a single incident probe angle. The elastic inhomogeneity was extracted from the analysis of the variations in the BF by neglecting the inhomogeneity of the optical reflective index. A subwavelength spatial resolution of 40 nm was achieved by direct fitting the BO in the time domain, thus avoiding a windowed Fourier transform of the signal.

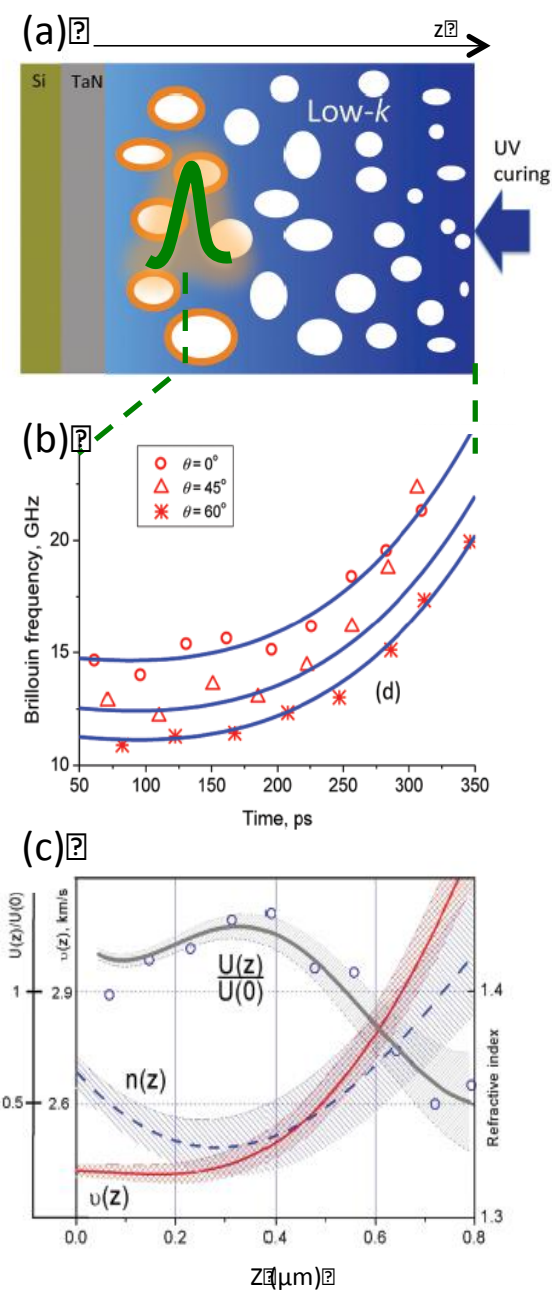

**FIG. 7**. First experiments on depth-profiling by TDBS of a continuously distributed in space elastic inhomogeneity of material. (a) Representation of the sub-micron inhomogeneities that exist within nanoporous transparent films. The white disks are nanopores. Their dimensions vary from the free surface toward the TaN - low-$k$ interface. The blue matrix is an organo-siloxane matrix. The color gradient indicates that the deeper (lighter) blue is the region of higher (smaller) elastic modulus. Some nanopores still contain polymeric porogen (orange edges of the pores) that was not completely removed during nanomanufacturing. For optical generation of the CAPs, a TaN layer of 30 nm thickness was deposited on a Si substrate before deposition of the low-$k$ material (Fig. 1a). (b) Time-dependent BF of the transient reflectivity signal obtained at different probe beam incidence angles $\theta$ [28]. (c) Depth profiles of the refractive index, $n(z)$, the sound velocity $\upsilon_s(z)$) and the normalized photoelastic coefficient, $P(z)/P(0) \approx U(z)/U(0)$, obtained by the PAI [28]. Reproduced with permission from ACS Nano **6**, 1410 (2012). Copyright 2012 American Chemical Society.

Experiments reported later[8] on the same samples were conducted in the so-called "red pump – blue probe" ($\propto 880$ nm – $\propto 440$ nm) configuration as presented in Fig. 1 (a). It was demonstrated[28] that by performing picosecond ultrasonic interferometry at a few angles $\theta$ of probe light incidence[87,88] it is straightforward to determine from the BF temporal variations, $\omega_B(t) = 2\pi\nu_B(t)$, both the profile of the refractive index $n(z)$ and the profile of the sound velocity $\upsilon_s(z)$ simultaneously. This opportunity is based in its essence on the dependence of the BF in non-collinear interactions of probe photons with the CAP in the backward BS configuration on the angle between the directions of the propagation of the coherent acoustic waves and the probe laser light, $\nu_B(t) = (1/\pi)\upsilon_s(z_a)k[n^2(z_a) - \sin^2(\theta)]^{1/2}$. Note, that this theoretical formula, implemented in Ref. 28 for depth-profiling, is also the basis for the interpretation of several other TDBS experiments conducted in the homogeneous materials[87-91]. In the inhomogeneous medium it is valid within the framework of geometrical optics[67,68], while $z_a = z_a(t)$ is the current position of the CAP. As the CAP with initial magnitude $\eta(0)$ and duration $\tau$ propagates along the $z$-axis from the transducer at $z = 0$ (Fig. 1 (a)) its strain magnitude and spatial scale vary (within the framework of the geometrical acoustic approximation[67,68]) as $\eta(t) = \eta(0)[\upsilon_s(z_a)/\upsilon_s(z)(0)]^{-3/2}[\rho(z_a)/\rho(0)]^{-1/2}$ and $\upsilon_s(z_a)\tau$, respectively, where $\rho(z)$ is the distribution of density. Moreover, because of the variations of the CAP amplitude, of its spatial extension, and of the photo-elastic constant $P(z_a)$ in depth, the BO amplitude is time-dependent: $A_B(t) \propto P(z_a)[n^2(z_a) - \sin^2(\theta)]^{-1/2}[\upsilon_s(z_a)\rho(z_a)]^{-1/2}$.

Figure 7 (b) gives $v_B(\theta,t)$ evaluated[28] at three different angles ($\theta = 0°$, 45° and 60°). The dependencies $\upsilon_s(t)$ and $n(t)$ extracted from $v_B(\theta,t)$ were transformed using the space-time correspondence, $z = z_a(t) = \int_0^t \upsilon_s(t')dt'$, into depth distributions of $\upsilon_s(z)$ and $n(z)$ (Fig. 7 (c)) and also were used together with the time-dependent amplitudes $A_B(t)$ to examine variations in space of the parameter $U(z) \propto P(z)\rho(z)^{-1/2}$. Experimentally determined, non-monotonic variations of the optical density $n$ along the depth coordinate z of the nanoporous low-k film was correlated[28] with its non-monotonic dependence on the UV radiation dose. The photoelastic coefficient *P(z)* was estimated to be the dominant contributor to the variations in *U(z)*. Overall, the experimental results presented in Fig. 7.(c) reveal a wealth of nanoscale subsurface information. An improved understanding of the depth-dependent UV curing effect is a key factor for successful integration of ultra-low-*k* layers in advanced microelectronic technology, while nondestructive depth-profiling of the mechanical, optical, and acousto-optical properties is essential for achieving this purpose.

D. Imaging of inhomogeneity and texture of polycrystalline materials at high pressures

TDBS was introduced for the first time in high pressure research in 2008[92,93]. However, its applications to depth-profiling and two-dimensional imaging of a polycrystalline aggregate of ice compressed in a diamond anvil cell to megabar pressures were reported only quite recently[47] (Fig. 8 (a)).

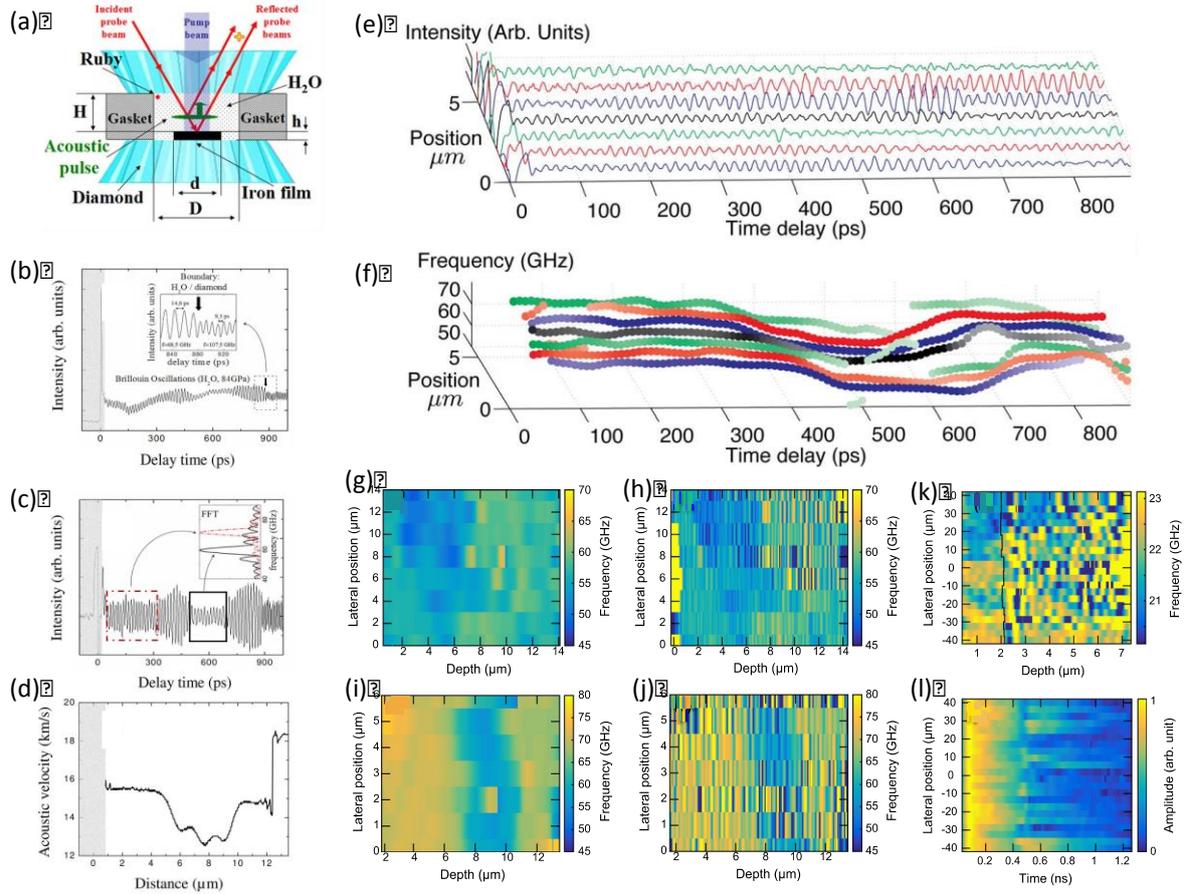

**FIG. 8.** Application of time-domain Brillouin scattering in high pressure research. (a) Transparent, spatially inhomogeneous medium (polycrystalline $H_2O$ ice[47] compressed in a diamond anvil cell. A picosecond pulse from the pump laser (violet) hits the absorbing transducer launching a coherent acoustic pulse (green) in the sample volume. Propagation of the pulse is monitored by a delayed probe laser pulse (red) reflected from stationary (transducer) and moving (acoustic pulse) interfaces. (b)-(c): Demonstration of the capability of TDBS for revealing spatial inhomogeneities in $H_2O$ ice X at megabar pressures. (b): Typical time-resolved reflectivity signal in ice compressed in a DAC to 84 GPa. The vertical arrow marks the time of transmission of the laser-generated coherent acoustic pulse across the interface of ice with diamond. Insert: Zoom of the signal in the vicinity of the ice/diamond interface. (c) TDBS signal obtained by subtracting from the signal in (b) the time-varying thermo-reflectance contribution caused by transient heating of the sample. Insert: Fourier spectra of the signal inside two temporal windows, marked by rectangles showing the shift of the Brillouin frequency from 60 GHz to approximately 73 GHz, indicating spatial inhomogeneity in the ice sample. (d): Spatial variation of the longitudinal sound velocity obtained from the temporal dependence of the BF, as revealed by a Fourier transform performed in a rectangular moving temporal window of 100 ps duration, using extrapolated refractive index values of ice[96]. (e) – (f): Demonstration of µm-scale texture in $H_2O$ ice aggregate at megabar pressures. (e): Intensity versus time delay signals in $H_2O$ ice at 84 GPa, obtained by displacing the sample relative to the co-focused pump and probe laser beams in the lateral direction, i.e., parallel to the ice/diamond interface in 1 µm steps. (f): Two-dimensional images of the Brillouin frequency magnitude obtained by processing the signals in (e). Isometric representation of the TDBS signals and of the temporal dependence of the Brillouin frequency of ice X. In (f) the signal amplitude is correlated with the symbol color: the darker the symbol, the higher the maximum amplitude of the BO, and vice versa. Image (f) reveals a clear, large-scale, 3–5µm, layering of the ice aggregate at 84 GPa in the direction normal to the diamond anvil culets. Reproduced with permission from Sci. Rep. **5**, 9352 (2015). Copyright 2015 Nature Publishing Group. (g) – (l): 2D-images of Brillouin frequency inhomogeneities in water ice compressed in a DAC to 57 GPa [(g) and (h)] and 84 GPa [(i) and (j)]. Absolute values of the frequencies can be derived from the color bars on the right hand side of each plot. The in-depth spatial resolution is about 1.05 µm in (g) and (i), and 0.26 µm in (h) and (j). (k) – (l): 2D-images of Brillouin frequency (k) and of the Brillouin oscillation amplitude (l) of solid argon compressed in a DAC to 15.4 GPa[48]. The in-depth spatial resolution is about 0.36 µm. In (k) the black line stands for delimiting the depth after which the influence of the second acoustic pulse on the images could be important. Reproduced with permission from Ultrasonics, 69, 201 (2016). Copyright 2016 Elsevier.

Knowledge of the pressure dependences of sound velocities and elastic moduli of liquids and solids at megabar pressures and the evolution of the texture of polycrystalline samples under compression is of extreme importance for a few branches of the natural sciences, such as condensed-matter physics, physics of the Earth and planetology, as well as for the monitoring and prediction of earthquakes and tsunamis, and for monitoring nuclear weapons tests. Experimental data on high-pressure materials parameters can be substantially influenced by spatial inhomogeneities and texture variations in polycrystalline aggregates. The aggregate sound velocity depends on the characteristic dimensions of individual crystallites, which are elastically anisotropic[94], and even a partial alignment of the crystallites i.e., orientational texture, which biases the measured velocities[95]. These factors prevent precise evaluation of elastic moduli and require the development of experimental methods for three-dimensional imaging of microscopic samples in situ when compressed in a diamond anvil cell (DAC) for the purpose of characterizing both their morphological and orientational/directional texture.

Figures 8 (b-d) illustrate how the inhomogeneity of $H_2O$ ice was revealed in TDBS experiments[47]. Figures 8 (e-j) present the images of the spatial structuring of ice. In Fig. 8 (g,h) the two-dimensional images of ice obtained with different spatial resolution reveal different scales in material inhomogeneity and texturing. In Fig. 8 (k) the 2D image of Ar ice in DAC[48] is presented for comparison with those of $H_2O$ ice revealing greater lateral than axial structuring in polycrystalline Ar. The most detailed updated information on the analysis of the TDBS imaging signals measured in ices[47,48,97] can be found in Ref. [97]. The synchronous detection technique (SDT) was applied in the temporal window equal to one Brillouin oscillation. The dependence of the accuracy in frequency determination on the signal-to-noise ratio (SNR) and the number of signal points inside the window was described. Two goodness criteria were applied in the signal analysis. The first one requires that the SNR exceeds 10 dB and the second one that the normalized root-mean-square error (NRMSE) remains below 10%. Under these criteria the frequency in the temporal window containing one Brillouin cycle was determined with the accuracy better than 6%.

The TDBS method allowed examination of the characteristic dimensions of ice texturing with sub-micrometer spatial resolution in the direction normal to the diamond anvil surfaces and with micrometer spatial resolution in the lateral directions. It is worth noting here that the axial spatial resolution in classic BS microscopy[14] applied to three-dimensional imaging[15] currently exceeds 2 micrometers. In chemically homogeneous transparent aggregates TDBS imaging[47,48] provides for each crystallite (or group of crystallites) usable information on its orientation (if the material is elastically anisotropic) as well as the elastic modulus along the direction of sound propagation. From the viewpoint of quantitative measurements the TDBS technique is significantly superior to the classical frequency-domain BS (FDBS) technique as a result of its far higher spatial resolution. For example in Ref. 47 the values of the longitudinal sound velocity, $\upsilon_s(z)$, in $H_2O$ ice were recorded with an axial resolution of ~0.26 $\mu m$ at 82 GPa, corresponding to the length of one BO. Thus, the TDBS signal collected at one given lateral position from a 15 $\mu m$ thick sample contained up to 60 independent $\upsilon_s$ measurements. In order to accumulate statistically reliable limits of variation of $\upsilon_s$ lateral scans over distances up to ~70 μm can be performed. This capability dramatically increases the statistical confidence levels of the recorded maximal and minimal velocities in cubic single crystals, $\upsilon_s^{<11\flat>}$ and $\upsilon_s^{<100>}$, permitting reliable determination of pressure dependences of single crystal elastic moduli $c_{ij}(P)$ of ice at high pressures[93].

E. Subcellular imaging inside plant and animal cells

Owing to its demonstrated capability for determining materials properties and imaging spatially localized inhomogeneities in solids, TDBS was explored for probing liquids[98,99].

After the first experiments in water[98] were reported, the interest shifted towards demonstration of the utility and biocompatibility of the technique for probing cells[79,80,99-103], the goal being development of a tool for imaging at a subcellular scale for biology and medicine. Different BFs were measured separately for the vacuole and nucleus *in vitro* in individual onion cells[79] suggesting that in principle there could be sufficient acoustic contrast within the cell for imaging. The estimated acoustic stress generated with TDBS is much lower than the adhesion stress of cells to their surroundings making the technique completely non-invasive[80,102].

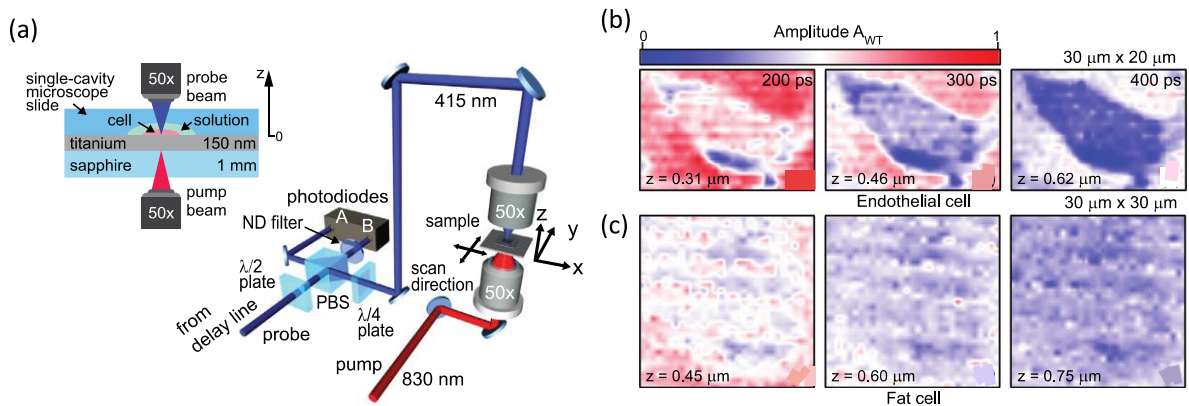

**FIG. 9.** TDBS imaging in cells in probe reflection configuration. **(a):** Experimental setup for cell imaging[49]. PBS: polarization beam splitter. ND filter: neutral density filter. Each (unstained) cell was covered by a single-cavity microscope slide filled with saline and buffer solutions for the endothelial and fat cells, respectively. Each slide is then placed on an x–y translation stage, allowing 2D raster scans. The probe beam is temporally scanned up to 800−1000 ps (after each ultrasonic pulse arrives at the Ti-solution interface) through use of a mechanical delay line. Spatial and temporal scanning allow 3D imaging (x, y, and t) over 40 μm× 40 μm areas with $\propto$1 μm and 1 ps resolutions. Two types of mammalian biological tissues—a bovine aortic endothelial cell and a mouse adipose, cell were investigated. Normalized wavelet transformed amplitude images $A_{WT}(x, y)$ at the Brillouin frequency at different times: (b) at a frequency of 10.3 GHz for the endothelial cell at 200, 300, and 400 ps and **(c)** at 9.9 GHz for the fat cell at 300, 400, and 500 ps. The calculated average propagation distances z of the ultrasonic pulses within the cells at these times are also shown. Reproduced from Appl. Phys. Lett. 106, 163701 (2015), with the permission of AIP Publishing.

Although the TDBS has been applied to the spatially resolved imaging of the cells only recently[49-51] the method has already shown promise. In Ref. 49 TDBS allowed three-dimensional imaging (x, y, and t) of the CAP propagation with $\propto 1$ $\mu m$ lateral and $\propto 150$ nm depth resolutions in vitro animal cells (Fig. 9 (a)). The acquisition time for a single image ($\propto$ 30 s) is of the same order as that in super-resolution microscopy techniques such as photoactivated localization microscopy (PALM) and stochastic optical reconstruction microscopy (STORM)[12,13]. The BOs were wavelet transformed using the Morlet mother wavelet. The spatial depth resolution was limited to the order of the Brillouin wavelength. The frequency of the oscillations was measured with the accuracy $\propto 1\%$. Spatially interpolated images of normalized wavelet transform amplitude (AWT) at three separate times are shown in Figs. 9 (b) for an endothelial cell and in 9 (c) for a fat cell. The spatial dependence of AWT(t) in Figs. 9 (b,c) shows evidence of inhomogeneity inside the cells. The application of PAI allowed[49] determination of important physical parameters of cell samples including sound velocities, bulk moduli, and ultrasonic attenuation coefficients.

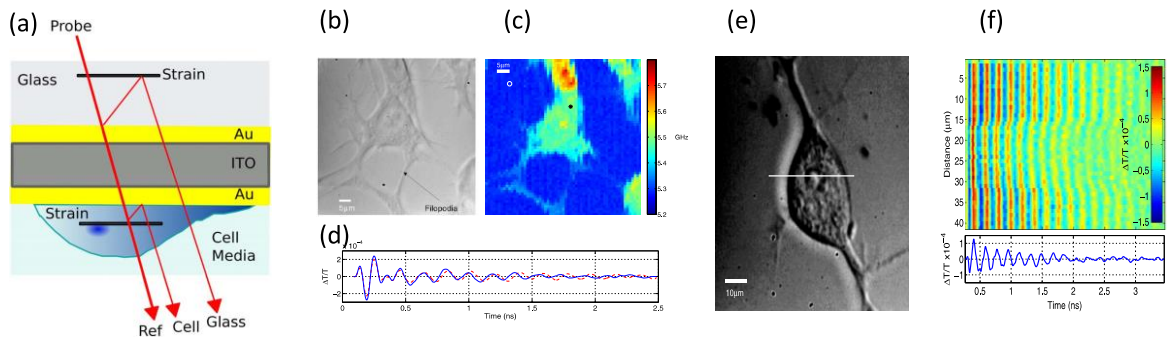

**FIG. 10.** TDBS imaging in cells in probe transmission configuration. (a): Schematic of the setup for operating in the transmission mode using a thin-film transducer[50]. Two signals are detected - one from the specimen (ref + cell) and one from the substrate (ref + glass). The dimensions of the transducer are exaggerated and beams are shown at an angle for clarity. 3T3 fibroblast cells were cultured on the transducers for 24 h using standard culture conditions prior to fixation in paraformaldehyde (4%) in phosphate buffered saline (PBS). The fixed cells were kept in PBS during the experiment, allowing a gentle flow that prevented air bubble formation. (b): Optical image of a 3T3 fixed cell, and (c), its TDBS map obtained using the transmittance approach. The scanned area was 60 by 60 μm, and was recorded with one-micron steps. (d): A representative trace obtained from the cell (blue) and the medium (dashed red), revealing possible spatial inhomogeneities in the axial direction, which, however, were not analyzed[50]. The locations of these traces are indicated by a star and a circle, respectively. (e)-(f): TDBS scanning over living 3T3 cells. A microscope image of the cell before the scan is given in (e). A B-scan of the acquired cross-section with a sample time trace taken from the 21 μm position is given in (f). The powers for pump and probe were 0.4 and 1 mw, respectively. The cell is exposed to 0.04 and 0.3 mW from the pump and probe beams owing to the transducer transmittance characteristics. The morphology of the cell after the scan remained structurally intact. Reproduced with permission from Applied Optics **54**, 8388 (2015) Copyright 2015 Optical Society of America.

The TDBS technique described in Ref. 50 used a three-layered metal-dielectric-metal film[104] as a transducer to launch acoustic waves into the cell (Fig. 10 (a)). The design of this transducer and measuring system was optimized to overcome the vulnerability of a cell to the exposure of laser light and heat without sacrificing the signal-to-noise ratio. The transducer substrate shields the cell from the laser radiation, efficiently generates acoustic waves, facilitates optical detection in transmission, and aids with heat dissipation away from the cell. The experiment[50] was based on a dual laser ASOPS system[83,84] .where the sample was scanned by moving electromechanical stages with a minimum step motion of 100 nm. In a single scan, typically 15,000 averages per point were taken, which took 4 s to acquire. Figures 10 (e-f) show two fixed 3T3 fibroblast cells and the result of a 2D scan of their BO. The correlation between features in the optical image (b) and the TDBS map (c) is good. It can be seen that the nuclei of both cells had clearly different Brillouin frequencies. This change is much less visible in the optical picture demonstrating the potential of acoustical contrast. Data acquisitions was also possible with living cultures of 3T3 cells as shown in Figs. 10 (e,f).

The experiments[49-51] reported to date suggest extension of TDBS techniques for high-resolution acoustic intercellular imaging which should prove to be invaluable for investigating the mechanical properties of cell organelles. The application of TDBS allows determination of important physical parameters including sound velocities, bulk moduli, and ultrasonic attenuation coefficients of animal[49,100] and vegetal[50,51,79,103] cells. Characterization of the physical properties of cells is essential for understanding mitosis, apoptosis, adhesion, and mobility.

## IV. IMAGING OF NANO-ACOUSTIC WAVE TRANSFORMATION IN HOMOGENEOUS MEDIA

### A. Attenuation of coherent hypersound

TDBS possesses a capability for following the transformation of CAPs during their propagation in optical transparent media as can be seen from examination of the basic formula for acoustically-induced transient reflectivity variations given in Eq. (1), where spatial-temporal distribution of the strain field enters under the integral. A well-known example of acoustic wave transformation is evaluation of coherent sound attenuation from the temporal decay of the Brillouin oscillation amplitude[23,25,92,98]. We cite here just some of the measurements of sound absorption in glasses[25], solids[92], polymers[105], glass-forming liquids[106], liquids[98,107], and biological cells[49,79]. It is important to note that the experimental results presented in Ref. 80 indicated deviation in the decay of the detected BO from an exponential one. The deviation was increasing with increasing propagation distance of laser-generated CAP from the emitting opto-acoustic transducer (OAT). The authors hypothesized that the revealed attenuation of the BO, additional to acoustic absorption, was caused the diffraction of the coherent acoustic waves as it had been expected theoretically[25]. This hypothesis was supported by the estimates of the characteristic diffraction length of the coherent acoustic wave at BF and is very plausible. The additional attenuation of the TDBS signal caused by the diffraction of the CAP was also reported later[108]. However, no attempts to image the CAP diffraction process by TDBS, via quantitative fitting of the BO attenuation, have been undertaken.

B. Nano-focusing of coherent hypersound

A very recent example is the application of TDBS for monitoring of the focusing of coherent nanoacoustic pulses[63]. Infrared pump light pulses were used to generate CAPs at the surface of a sub-micron, single metal-coated silica fibre (Fig. 11 (a)). Partially transmitted probe light pulses allowed the CAPs to be continuously monitored revealing acoustic focusing in the time domain as a result of the curvature of the fibre surface. An analytical model, supported by three-dimensional simulations, suggested that focusing of the 44-GHz BF of the acoustic beam was possible down to a ~ 150-nm diameter waist (Fig. 11 (b)). The experimental results, Fig. 11 (c), are in good agreement with the theoretical predictions.

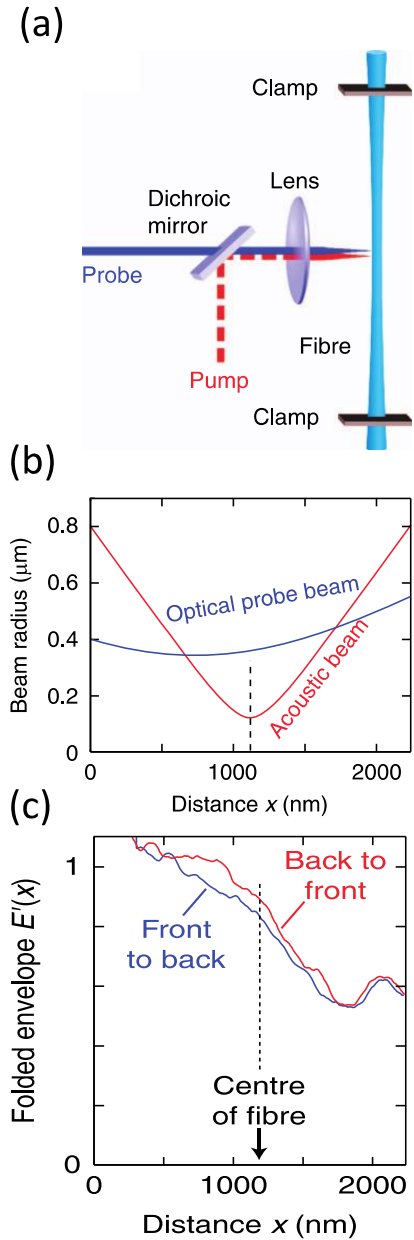

FIG. 11. Optical tracking of picosecond coherent phonon pulse focusing inside a sub-micron object[63]. (a): Experimental sample and setup. (b): Theoretical optical probe (blue) and acoustic-strain (red) $1/e^2$ beam radii as a function of propagation distance x (over the region from the front surface of the fibre to the back) as determined from Gaussian-beam theory for a fibre of radius R=1150 nm. (The dashed line indicates the position of the centre of the fibre.) (c): Experimental results for the BO envelope as a function of the propagation distance x. Reproduced with permission from Light: Science & Applications **5**, e16082 (2016). Copyright 2016 Nature Publishing Group.

The work with fibre generated CAPs demonstrates the high lateral resolution of TDBS imaging possible with focused picosecond CAPs, which, otherwise, is normally limited by the diffraction limit of the pump optical pulses to ~ 1 μm. In particular, until now TDBS imaging has not been able to overcome the optical diffraction limit for lateral resolution. As suggested in Ref. 63, with the use of metal-coated cylindrical or spherical sub-micron hollowed regions on the end of an immersed tapered fibre in a standard near-field scanning optical microscope geometry, TDBS could follow focused GHz acoustic phonon pulses in liquids, extending the resolution of 3D acoustic imaging[49-51] of biological structures to the nanoscale. In addition to focusing of CAPs[63,109] other possibilities for improving the lateral resolution of TDBS imaging include photo-generation of CAPs significantly narrower than the pump laser beam dimensions as a result of nonlinear processes in opto-acoustic transformation[110] and application of near-field optical schemes[111,112].

C. Nonlinear transformation of coherent nano-acoustic pulses

Variation of the CAP spatial-temporal distribution in homogeneous media can also be caused by nonlinear acoustic phenomena, which are known to lead to the formation of weak shock fronts in non-dispersive media[113]. Frequency mixing processes arising from the elastic nonlinearity of the material modify the spectrum of the CAP, its shape and length. The nonlinear transformation of bipolar picosecond CAPs had been first observed by classical frequency-domain BS[114] through measurement of the distribution of 22 GHz coherent longitudinal phonons along the CAP trajectory via a step-wise increase of distance between the opto-acoustic transducer and a continuous probe light focusing region. The resolution in the propagation direction, on the order of 4μm, was determined by the waist of the focused probe laser. This spatial resolution is 1 – 2 orders of magnitude lower than that achievable by TDBS. Reference 61 describes experiments where nonlinear bipolar CAPs propagating in a SrTiO$_3$ (STO) substrate were monitored by TDBS over a range of optical wavelengths from 700 nm to 470 nm. Beats in the amplitude of the detected BO were observed at high laser fluences (Fig. 12 (a)).

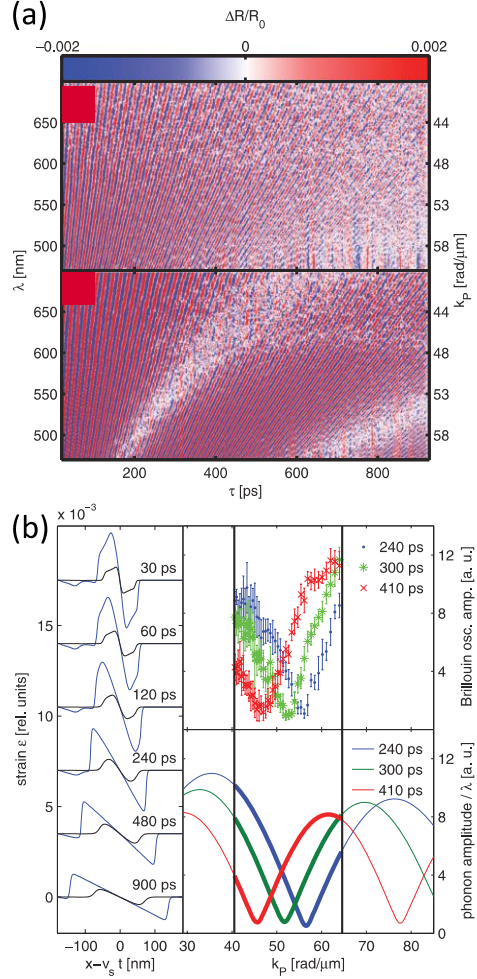

**FIG. 12.** Nonlinear transformation of a bipolar coherent acoustic pulse as followed by time-domain Brillouin scattering[61]. (a): Relative optical reflectivity change of the sample as a function of delay time for a pump fluence of 14 mJ/cm$^2$ (upper panel) and 47 mJ/cm$^2$ (lower panel). The low-frequency background was subtracted from the image by high pass filtering. The ordinate gives the probe pulse wavelength $\lambda$ (left axis) and the wavenumber of a phonon $k_p$ causing Brillouin scattering (right axis). For strong excitation conditions (lower panel) a beating was observed in the TDBS signals. (b): (left panel) Spatial profile of the bipolar strain pulse in the STO for different propagation times in a frame of reference propagating with the speed of sound $v_s$ for high amplitude (blue line, 0.47% strain) and low amplitude (black line, 0.14% strain). (upper right panel): Measured amplitude of the Brillouin oscillation for each $k_p$. The region between the vertical black lines indicates the wave vectors that can be accessed by the optical white light. (lower right panel) Phonon amplitude divided by the wavelength $\lambda$ as a function of wave vector calculated from Fourier transforms of the simulated strain profile. The data show good agreement with the measurement in upper panel. Reproduced with permission from Phys. Rev. B **86**, 144306 (2012). Copyright 2012 American Physical Society.

Theory[61,69] indicates that the minima in the BS spectrum at a certain time delay at particular frequencies (Fig. 12 (b) right panel) can be interpreted as mutual compensation of the probe light fields scattered by leading and trailing fronts of the bipolar CAP. Correspondingly, the observed beats (Fig. 12 (a)) are a consequence of the constructive and destructive interference of the probe light scattered by two acoustic fronts moving relative to each other[61,69].

Monitoring of the nonlinear evolution of the unipolar CAPs by TDBS was achieved in Ref. 62. A unipolar CAP theoretically[69,113] transforms in the time-domain, as indicated in the upper insert of Fig. 13 (a), to a triangular form with a weak shock front and a trailing edge of far longer duration. In the frequency domain nonlinear phenomenon of frequency-mixing leads to continuous generation and absorption of phonons at the BF accompanied by the phase shift. In the time-domain, continuous generation and absorption processes result in the BF predominantly localized at a weak shock front. Consequently, the velocity of weak shock front detected by TDBS matches the velocity of BF component[69]. Thus, the difference $\partial \varphi_\eta / \partial t$ between the cyclic frequency detected by TDBS $\omega_{dR/R}$ and the linear BF $\omega_B$ Eq. (3), is due to the additional nonlinear transport of the BF phonons and other high frequency phonons contributing to weak shock front. The shock speed, and, as a consequence $v_{dR/R}$ increase with CAP amplitude, which increases with laser fluence and diminishes with the propagation distance, as was observed experimentally in Ref. 62 (Figs. 13 (a,b)) in accordance with theoretical expectations[69]. Non-exponential reduction of the frequency shift $v_{dR/R} - v_B$ with time shown in Fig. 13 (b) for high excitation fluences has also an origin in nonlinear acoustics[62,69]. It is worth noting here that in the experiments[62] the duration of the acoustic pulse initially launched in the medium was approximately ten times shorter than the period of the Brillouin oscillation, $T_B$, but could become comparable to $T_B$ at the largest imaged depths for the highest pump laser fluencies. At the same time for the highest laser fluencies the width of the shock front, which is controlled by the competition of the nonlinearity and dissipation, is shorter than the initial duration of the acoustic pulse even at the end of the experimental time window, i.e., at largest depths. The nonlinear dissipation of acoustic energy[69,113] takes place in the weak shock front and the points of the shock front profile are the only points of the CAP profile, which are moving at time-dependent velocity. Thus, revealing by the TDBS of the time-dependent $\omega_{dR/R}$[62] provides an experimental evidence of the TDBS sensitivity to the processes taking place in the shock front and of the TDBS imaging with spatial resolution much higher than one limited by the Brillouin wavelength, $\lambda_B$. TDBS enabled[62] determination of the acoustic nonlinearity parameter[113] to which the wave-amplitude-dependent quantity $\partial \varphi_\eta / \partial t$ is proportional[69] (Fig. 13 (c)). A material-specific nonlinear acoustic parameter in a solid characterizes the anharmonicity of the intermolecular interaction potentials; in a liquid the same parameter should reflect an additional nonlinearity arising from structural relaxation. The experiments reported in Refs. 61 and 62 demonstrate convincingly that TDBS opens the door to versatile measurements of gigahertz nonlinear acoustic phenomena.

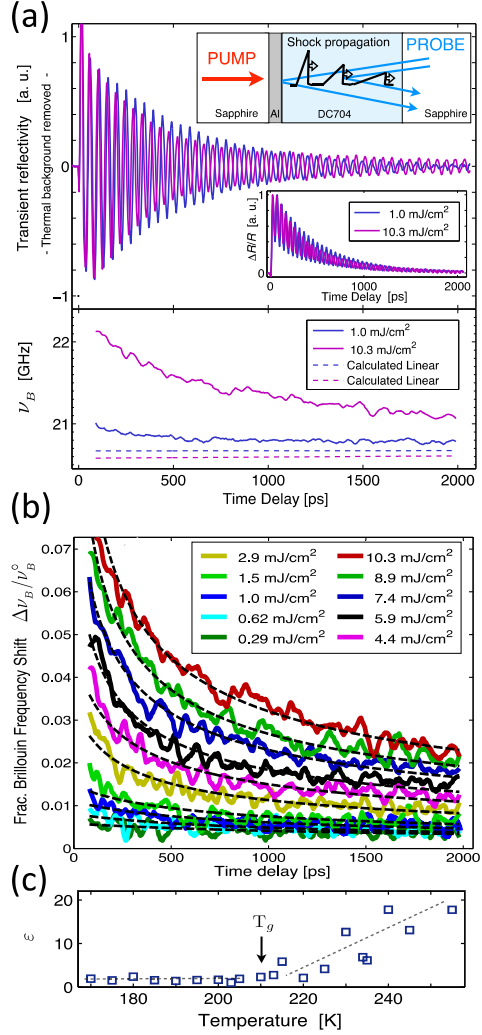

**FIG. 13.** Nonlinear transformation of a unipolar coherent acoustic pulse as followed by time-domain Brillouin scattering[62]. (a): (upper panel): Normalized TDBS data recorded at 200 K in silicone oil (DC704) at two representative low and high pump fluences with the thermoreflectance background removed[62]. Inset (top): Experimental setup with the ~100 μm liquid squeezed between two sapphire substrates, one holding a 33 nm aluminum photoacoustic transducer film. A propagating nonlinear unipolar CAP is depicted as well. Inset (bottom): Diagram of the nonoscillatory signal component arising from thermally induced changes in reflectivity of the Al film. (lower panel) Extracted values for the time variation of the experimentally determined Brillouin frequency. The frequency-down chirp with increasing time at high fluence is caused by nonlinear acoustic propagation of a weak shock. The dashed lines show the calculated up-chirp due to sample heating at 1.0 and 10.3 mJ/cm$^2$ pump fluence (top and bottom lines, respectively). (b): Measured results of the fractional Brillouin frequency shift, ($\nu_{dR/R}$ -$\nu_B$ ) / $\nu_B$ in DC704 for different laser pump fluences. Dashed lines are theoretical fits based on an analytical solution for nonlinear transformation of a triangular CAP into a weak shock. (c): The dependence of the nonlinear acoustic parameter of the fragile glass former DC704 at around 20 GHz at temperatures extending through the glass transition. Reproduced with permission from Phys. Rev. Lett. **114**, 065701 (2015). Copyright 2015 American Physical Society.

## V. PERSPECTIVES AND CONCLUSIONS

It is evident from the extensive experimental advances seen to date that the possibilities for application of TDBS to nanoscale imaging are numerous. There exist some limitations and drawbacks of the TDBS scattering technique to imaging and some prospects in enhancing its functionalities. As it have been explained in Section II the spatial depth resolution of the TDBS imaging can be significantly better than one limited by the wavelength of the acoustic phonon at the BF, $\lambda_B$. In all currently known TDBS experiments the duration of the laser pulses and the time interval $\Delta t$ between neighbor points in the signal are shorter than the duration of CAP, so the length of the CAP or of its sharp front localization are expected to be the limiting factors in the spatial resolution. However the spatial resolution was limited by the duration of the moving window, $\tau_w$, in which the theoretical signal was fitted to the experimental one, and in most of the experiments the window, $\tau_w$, was not narrower than the Brillouin oscillation period, $T_B = 2\pi / \omega_D$. Thus in most of the TDBS imaging experiments the spatial resolution was not better than $\lambda_B$. There could be several possible reasons for choosing $\tau_w \geq T_B$. First, in this large window a quasi-sinusoidal character of the signal is well determined and the well-developed/known approaches based on the Fourier and wavelet transforms could be applied. Second, in the case of the probe laser radiation at wavelength $\lambda_p \propto 400$ nm the Brillouin wavelength, $\lambda_B$, is around 100 nm, depending on the optical refractive index, and this spatial resolution could be sufficient for the experimental goals. Third, the shortening of $\tau_w$ additionally 4-5 times to a typical duration of the CAP could require improvement/modification of the experimental setup and would require application of less common/known methods for signal fitting. In the inhomogeneous medium in general both the amplitude and the phase of the TDBS signal are varying and each of them has to be approximated by a polynomial, for example, increasing the number of the fitting parameters. Then the minimum duration of $\tau_w$, which could be achieved, in general depends on the number $N \approx \tau_w / \Delta t$ of the experimental points in the moving window, on the signal-to-noise ratio (SNR) and on a desired accuracy in the determination of the parameters. The larger is SNR and $N$ the smaller could be uncertainty in the evaluation of the parameters. Shortening of $\tau_w$ without lose in accuracy could require increase of $N$ through the diminishing of $\Delta t$ and the increase of SNR. Both could be achieved by increasing the experimental time if it is not limited by the stability of the laser or of the sample. When limited by the available experimental time the SNR could be still increased by longer acquisition time in each experimental point and $\Delta t$ could be diminished conserving the total duration of the experiment but diminishing the imaged depth of the sample.

It is also obvious that to achieve the ultimate spatial resolution would be more and more complicated if the number of CAPs propagating simultaneously in an optically transparent sample increases, because of the necessary increase in the number of the fitting parameters. For example, this situation could take place when thin film OAT in Figs. 1 (a) is not well matched acoustically to the substrate or the tested inhomogeneous medium and emits a sequence of CAPs comparable in amplitude and periodically separated in time by the interval $\tau_d$ equal to the time required for the CAP to cross the film twice[25,48,69]. If it is impossible to avoid this type of the OAT ringing, then it could be worth trying to accomplish first the extraction of the distributions of material parameters from the data accumulated in the time window $0 \leq t \leq \tau_d$ and then to use the obtained results for theoretical prediction of the contribution to TDBS signal from the second CAP when $\tau_d \leq t \leq 2\tau_d$ and its deletion from the total TDBS signal. Thus the imaging by a single first CAP could be extended to the interval $\tau_d \leq t \leq 2\tau_d$, and so on… However to keep the necessary level of accuracy in determining the material parameters could require broadening of $\tau_w$ to compensate the uncertainties accumulating at each next time interval. Similar approach could be suggested if a single CAP initially travelling in a sample composed of several inhomogeneous layers of different transparent materials splits at the interface between the layers into two CAPs (reflected and transmitted) providing comparable contributions to the TDBS. However, in the latter case, if the BFs in the two simultaneously probed layers are sufficiently different, it could be more profitable to fit the TDBS signal from two CAPs simultaneously, by extending the theoretical formulas and increasing the number of the fitting parameters. In any case, it is obvious that for achieving the higher spatial depth resolution it is highly desirable to realize in the TDBS experiments the conditions for imaging by a single CAP at a time.

The limitation on the maximum depth of TDBS imaging could be caused by the attenuation of the CAP and/or of the probe light, and by decoherence of the reflected beams[25]. The probe light pulses scattered by the CAP should overlap in time in order to interfere at the photodetector. For example, in the simplest situation depictured in Fig. 3, if TDBS signal does not disappear at shorter distances due to limited penetration depths of CAPs and/or probe light, it will disappear at depths of CAP penetration exceeding half of the probe laser pulse length in the medium. For typical laser pulses of 200 fs duration these depths are $\geq 10$ $\mu m$ for the refractive indexes $n \leq 3$. This limitation is rather technical than physical. If required, it could be lifted off by using longer laser pulses. However, if the durations of the probe and pump laser pulses are mutually related and comparable, then possible elongation of the laser pulses is itself limited, because it could cause elongation of the CAP[36,37]. Another way to avoid the considered limitation could be splitting of a part of the probe beam before its incidence on the sample, delaying this part in time and using it for the heterodyning of the probe light scattered by the CAP.

From the technical point of view, new developments in TDBS imaging can be expected due to possible acceleration in signal acquisition, which should provide faster access to larger amounts of data than is presently possible (e.g. to a larger number of scans and more pixels per image), improvement in the signal-to-noise ratio through increased averaging at a particular measurement point, and a decrease in maximum thermal and optical loadings of the samples. The ASOPs technique[73,83,84] has already been applied for TDBS imaging[50-51], while multichannel[115] and wide-field[116] systems have not yet seen experimental application. Also from the technical point of view, samples for TDBS imaging could be prepared in such a way that most of the spurious signals degrading the TDBS signal, such as contributions to transient reflectivity caused by photo-excitation of the optoacoustic transducer, are suppressed. For example, it would be advantageous to have spatial separation between the regions where the CAP's are generated and the points where the inhomogeneities lie, as was done in some earlier picosecond ultrasonics experiments[117,118].

From the technical-physical point of view the prospects are related to the application in TDBS imaging of ultrafast time-resolved optical interferometry[119,116] and optical polarimetry/ellipsometry[33,121-126]. Theory[65] shows the advantages of depth profiling that could be gained if measurements of both the real and the imaginary parts of the transient reflectivity $dr(t)/r$ are determined by optical interferometry. In the ideal case, ultrafast interferometry presents the opportunity to separate exactly the slowly varying amplitude from the equally slowly varying frequency of the BO. It is worth noting here that the separate measurements of the phase and of the amplitude of the TDBS signal by optical interferometry will require corresponding modifications of the current methods applied to TDBS signal processing. For example, all the methods exploiting somehow the Fourier transform become senseless, because neither amplitude nor phase is periodic in time. The explicit theoretical solutions for local BF as a function on time derivative of the measured phase of the TDBS signal could be applied for $\omega_B(t)$ extraction[65]. Thus, the achievement of the ultimate spatial resolution would be related to the minimization of the temporal window in which the phase (or phase derivative over time) could be adequately fitted to provide the required accuracy in the determination of $\omega_B(t)$ when substituted in the analytical formulas. However, in most of the experiments this approach for absolute measurements of $\omega_B(t)$ requires additional independent measurement of some optical parameters of the samples and dedicated reduction/elimination of non-acoustic contributions to the TDBS signals[65]. For some cases, where the non-acoustic components masking the "wanted" TDBS signal are present, theory proposes how multiple measurements at different angles and/or with different probe polarizations could be applied in order to extract the wanted variations of the amplitude and frequency. These prospects are not simply technical because the application in TDBS of the polarimetry, for example, provides access to imaging of the material parameters that are not accessible by common TDBS technique, such as optical refractive index measurements for ordinary and extraordinary light waves and determination of the different components of the photoelastic tensor, including those controlling acousto-optic mode-conversion of probe light[126]. By conducting TDBS at several polarizations of the probe light it has already been possible to reveal spatially inhomogeneous optical anisotropy induced by loading of the material in a DAC[47] and to characterize spatial distributions of different photoelastic parameters of GaAs implanted by energetic protons[127]. Significant physical-technical progress also could be expected through the application of shear CAPs for TDBS imaging. Although optoacoustic transducers for shear CAPs are more elaborate than those used for compression/dilatation waves[33,121-133] and detection of shear CAPs as well[33,122,133], their application would provide access to imaging of the complex shear rigidity of inhomogeneous media at nanoscale.

From the physics point of view applications of TDBS imaging for deeper understanding of physical phenomena present themselves not only in the systems reviewed above but also in other materials, media, and structures that are spatially inhomogeneous at nanoscale. TDBS could be applied, for instance, to investigation of the physics of inhomogeneity of various transparent films, not simply those where the inhomogeneity is caused by post-deposition processing. The variation in the film texture with the increasing film thickness is considered a common phenomenon for film growth in general[134]. An inhomogeneity can also be introduced intentionally for the production of multilayered optical antireflective and highly reflective coatings[135]. Another possibility for the extension of physical applications of the TDBS technique is related to depth-profiling of the stress distribution near the surface of a crystal[136]. Knowledge of this distribution is not only of fundamental importance for surface and interfaces science but is also essential in microelectronics fabrication technologies. Stress in thin films in large scale integrated circuits can result in delamination of the film from the substrate, and cracking or buckling of the film itself. There can also be significant stress in the substrate as a result of the manufacturing process. Stress can come from the temperature gradients generated during solidification or as a result of deposition of structures on the surface. Thus, TDBS imaging of stress distributions in space could become an important technique for noncontact and nondestructive testing and control in micro and nano technology. Finally, the proven ability of the TDBS to monitor subsurface materials modifications in time[137,138], opens the path for the development of 3D spatial imaging of transient processes at nanoscale.

**Acknowledgements**

V.E.G. acknowledges support from the Alexander von Humboldt Foundation, JSPS, IUF and ANR grant LUDACism. The authors thank A. Lomonosov, O. B. Wright, T. Dekorsy, C.-K. Sun, A. V. Akimov, A. J. Kent, A. Zerr, C. Mechri, G. Vaudel, N. Chigarev, S. Raetz, A. Bulou, V. Tournat, V. Juvé and V. Temnov for collaborations and discussions. The authors thank G. J. Diebold and J. Thoen for critical reading and editing of the manuscript.

**References**

1. *Scanning Probe Microscopy and Spectroscopy: Theory, Techniques, and Applications*, Edited by D. Bonnell (2nd edition) (Wiley-VCH Inc., USA, ISBN 0-471-24824-X, 2000).

2. H.-K. Wickramasinghe, Scanned-probe microscopes, Scientific American **261**, 98-105 (1989).

3 C. B. Prater, H. J. Butt, P. K. Hansmaa, Atomic force microscopy, Nature **345**, 839-840 (1990).

4. D. Rugar, H. J. Mamin, P. Guenther, et al. Magnetic force microscopy: General principles and application to longitudinal recording media. J. Appl. Phys**. 68**, 1169 (1990).

5. P. Guthner, K. Dransfeld, Local poling of ferroelectric polymers by scanning force microscopy. App. Phys. Lett. **61**, 1137-1139 (1992).

6. G. Binnig, H. Rohrer, Ch. Gerber, and E. Weibel, Surface Studies by Scanning Tunneling Microscopy, Phys. Rev. Lett. **49**, 57 (1982).

7. A. Lewis, H. Taha, A. Strinkovski, A. Manevitch, A. Khatchatouriants, R. Dekhter and E. Ammann, Near-field optics: from subwavelength illumination to nanometric shadowing, Nature Biotechnology **21**, 1378 - 1386 (2003).

8. B. Bhushan, J. N. Israelachvili, U. Landman, Nanotribology: friction, wear and lubrication at the atomic scale. Nature **374**, 607 (1995).

9. Y. F. Dufrene, T. Ando, R. Garcia, D. Alsteens, D. Martinez-Martin, A. Engel, C. Gerber, and D. J. Muller, Imaging modes of atomic force microscopy for application in molecular and cell biology, Nature Nanotech. **12**, 295-307 (2017).

10. M. A. O'Keefe, P. R. Buseck and S. Iijima, Computed crystal structure images for high resolution electron microscopy, Nature **274** (5669), 322-324. (1978).

11. *X-ray and Neutron Refectivity: Principles and Applications* (J. Daillant and A. Gibaud, Eds.) Lecture Notes in Physics 770 (Springer, Heidelberg, 2009).

12. S. T. Hess, T. P. K. Girirajan, and M. D. Mason, Ultra-high resolution imaging by fluorescence photoactivation localization microscopy, Biophys. J. 91, 4258 (2006).

13. M. J. Rust, M. Bates, and X. Zhuang, Sub-diffraction-limit imaging by stochastic optical reconstruction microscopy (STORM), Nat. Methods **3**, 793–795 (2006).

14. K. J. Koski and J. L. Yarger, Brillouin imaging. Appl. Phys. Lett. **87**, 061903 (2005).

15. G. Scarcelli, W. J. Polacheck, H. T. Nia, K. Patel, A. J. Grodzinsky, R. D. Kamm and S. H. Yun**,** Noncontact three-dimensional mapping of intracellular hydromechanical properties by Brillouin microscopy, Nature Methods **12**, 1132 (2015).

16. C. Kak and M. Slaney, *Principles of Computerized Tomographic Imaging* (IEEE Press, New York,1988).

17. J. Krautkrämer, H. Krautkrämer, *Ultrasonic Testing of Materials*, 4th edition (Springer-Verlag, Berlin, 1990).

18. J. Blitz and G. Simpson, *Ultrasonic Methods of Non-destructive Testing*, 1st edition (Chapman & Hall, London, 1996).

19. C. Hellier, Ultrasonic Testing, Chapter 7 in Handbook of Nondestructive Evaluation (McGraw-Hill, New York, 2003).

20. R. S. Cobbold, *Foundations of Biomedical Ultrasound* (Oxford University Press, Oxford, 2007).

21. C. Thomsen, J. Strait, Z. Vardeny, H.J. Maris, J. Tauc, J. Hauser, Coherent phonon generation and detection by picosecond light pulses, Phys. Rev. Lett. **53**, 989 (1984).

22. C. Thomsen, H.T. Graham, H.J. Maris, J. Tauc, Surface generation and detection of phonons by picosecond light pulses, Phys. Rev. B **34**, 4129 (1986).

23. C. Thomsen, H.T. Graham, H.J. Maris, J. Tauc, Picosecond interferometric technique for study of phonons in the Brillouin frequency range, Optics Communications **60**, 55 (1986).

24. H.T. Graham, H.J. Maris, J. Tauc, Picosecond ultrasonics, IEEE J. Quantum Electron. **25**, 2562 (1989)

25. H.-N. Lin, R. J. Stoner, H. J. Maris, and J. Tauc, Phonon attenuation and velocity measurements in transparent materials by picosecond acoustic interferometry, J. Appl. Phys. **69**, 3816 (1991).

26. O. B. Wright, T. Hyoguchi, Ultrafast vibration and laser acoustics in thin transparent films, Optics Lett. **16**, 1529 (1991).

27. O. B. Wright, Thickness and sound velocity measurement in thin transparent films with laser picosecond acoustics, J. Appl. Phys. **71**, 1617 (1992).

28. A. M. Lomonosov, A. Ayouch, P. Ruello, G. Vaudel, M. R. Baklanov, P. Verdonck, L. Zhao, V. E. Gusev, Nanoscale noncontact subsurface investigations of mechanical and optical properties of nanoporous low-k material thin film, ACS Nano **6**, 1410 (2012).

29. K. E. O'Hara, X. Hu, and D. G. Cahill, Characterization of nanostructured metal films by picosecond acoustics and interferometry, J. Appl. Phys. **90**, 4852 (2001).

30. A. Devos, R. Côte, Strong oscillations detected by picosecond ultrasonics in silicon: evidence for an electronic-structure effect, Phys. Rev. B **70**, 125208 (2004).

31. A. Devos, R. Côte, G. Caruyer, and A. Lefèvre, A different way of performing picosecond ultrasonic measurements in thin transparent films based on laser-wavelength effects, Appl. Phys. Lett. 86, 211903 (2005).

32. I-J. Chen, P.-A. Mante, C.-K. Chang, S.-C. Yang, H.-Y. Chen, Y.-R. Huang, L.-C. Chen, K.-H. Chen, V. Gusev, C.-K. Sun, Graphene-to-substrate energy transfer through out-of-plane longitudinal acoustic phonons, Nano Lett. **14**, 1317 (2014).

33. D. H. Hurley, O. B. Wright, O. Matsuda, V. E. Gusev, O. V. Kolosov, Laser picosecond acoustics in isotropic and anisotropic materials, Ultrasonics **38**, 470 (2000).

34. C. Rossignol, B. Perrin, S. Laborde, L. Vandenbulcke, M. I. De Barros, and P. Djemia, Nondestructive evaluation of micrometric diamond films with an interferometric picosecond ultrasonics technique, J. Appl. Phys. **95**, 4157 (2004).

35. P. Babilotte, P. Ruello, D. Mounier, T. Pezeril, G. Vaudel, M. Edely, J.-M. Breteau, V. Gusev, and K. Blary, Phys. Rev. B **81**, 245207 (2010).

36. V. Gusev and A. Karabutov, *Laser Optoacoustics* (AIP, New York, 1993)

37. S. A. Akhmanov and V. Gusev, Laser excitation of ultrashort acoustic pulses : new advantages in solid state spectroscopy, the investigation of fast processes and nonlinear acoustics, Sov. Phys. Uspekhy **35**, 153 (1992).

38. P. Ruello, V. Gusev, Physical mechanisms of coherent acoustic phonons generation by ultrafast laser action, Ultrasonics **56**, 21 (2015).

39. J. F. Nye, *Physical Properties of Crystals* (Oxford Univeristy Press, Oxford, 1957).

40. P. M. Morse, *Theoretical Acoustics* (Princeton University Press, Princeton, 1968).

41. C. Mechri, P. Ruello, J.M. Breteau, M.R. Baklanov, P. Verdonck, V. Gusev, Depthprofiling of elastic inhomogeneities in transparent nanoporous low-k materials by picosecond ultrasonic interferometry, Appl. Phys. Lett. **95**, 091907 (2009).

42. A. Steigerwald, Y. Xu, J. Qi, J. Gregory, X. Liu, J.K. Furdyna, K. Varga, A.B. Hmelo, G. Lüpke, L.C. Feldman, N. Tolk, Semiconductor point defect concentration profiles measured using coherent acoustic phonon waves, Appl. Phys. Lett. **94**, 111910 (2009).

43. A. Steigerwald, A. B. Hmelo, K. Varga, L. C. Feldman, and N. Tolk, Determination of optical damage cross-sections and volumes surrounding ion bombardment tracks in GaAs using coherent acoustic phonon spectroscopy, J. Appl. Phys. **112**, 013514 (2012).

44. J. Gregory, A. Steigerwald, H. Takahashi, A. Hmelo, and N. Tolk, Ion implantation induced modification of optical properties in single-crystal diamond studied by coherent acoustic phonon spectroscopy, Appl. Phys. Lett. **101**, 181904 (2012); Erratum: Appl. Phys. Lett. **103**, 049904 (2013).

45. D. Yarotski, E. Fu, L. Yan, Q. Jia, Y. Wang, A. J. Taylor, B. P. Uberuaga, Characterisation of irradiation damage distribution near $TiO_2$/$SrTiO_3$ interfaces using coherent acoustic phonon spectroscopy, Appl. Phys. Lett. **100**, 251603 (2012).

46. A. Baydin, H. Krzyzanowska, M. Dhanunjaya, S. V. S. N. Rao, J. L. Davidson, L. C. Feldman, and N. H. Tolk, Depth dependent modification of optical constants arising from H+ implantation in n type 4H-SiC measured using coherent acoustic phonons, APL Photonics **1**, 036102 (2016).

47. S. M. Nikitin, N. Chigarev, V. Tournat, A. Bulou, D. Gasteau, B. Castagnede, A. Zerr, V.E. Gusev, Revealing sub-mm and mm-scale textures in H2O ice at megabar pressures by time-domain Brillouin scattering, Sci. Rep. **5**, 9352 (2015).

48. M. Kuriakose, S. Raetz, N. Chigarev, S. M. Nikitin, A. Bulou, D. Gasteau, V. Tournat, B. Castagnede, A. Zerr, and V. E. Gusev, Picosecond laser ultrasonics for imaging of transparent polycrystalline materials compressed to Megabar pressures, Ultrasonics, **69**, 201 (2016).

49. S. Danworaphong, M. Tomoda, Y. Matsumoto, O. Matsuda, T. Ohashi, H. Watanabe, M. Nagayama, K. Gohara, P. H. Otsuka, and O. B. Wright, Three-dimensional imaging of biological cells with picosecond ultrasonics, Appl. Phys. Lett. **106**, 163701 (2015).

50. F. Pérez-Cota, R. J. Smith, E. Moradi, L. Marques, K. F. Webb, and M. Clark, Thin-film optoacoustic transducers for subcellular Brillouin oscillation imaging of individual biological cells, Applied Optics **54**, 8388 (2015).

51. F. Pérez-Cota, R. J. Smith, E. Moradi, L. Marques, K. F. Webb, and M. Clark, High resolution 3D imaging of living cells with sub-optical wavelength phonons, Sci. Rep. **6**, 39326 (2016).

52. I. Chaban, H. D. Shin, C. Klieber, R. Busselez, V. E. Gusev, K. A. Nelson, and T. Pezeril, Time-domain Brillouin scattering for the determination of laser-induced temperature gradients in liquids, Review of Scientific Instruments **88**, 074904 (2017).

53. X. Wang, Y. Pang, G. Ku[1], X. Xie, G. Stoica and L. V. Wang, Noninvasive laser-induced photoacoustic tomography for structural and functional *in vivo* imaging of the brain, Nature Biotechnol. **21,** 803 (2003).

54. V. Ntziachristos, J. Ripoll, L. V. Wang, L. V and R. Weissleder, Looking and listening to light: the evolution of whole-body photonic imaging, Nature Biotechnol. **23,** 313 (2005).

55. T. Khamapirad, A. Conjusteau, M. H. Leonard, R. Lacewell, K. Mehta, T. Miller, A. A. Oraevsky, Laser optoacoustic imaging system for detection of breast cancer, *J. Biomed. Opt.* **14**, 024007 (2009).

56. *Photoacoustic Imaging and Spectroscopy*, Edited by Lihong V. Wang (CRC Press, 2009).

57. D. Razansky, M; Distel, C; Vinegoni, R. Ma, N. Perrimon, R. W. Köster and V. Ntziachristos, Multispectral opto-acoustic tomography of deep-seated fluorescent proteins *in vivo*, Nature Photonics **3,** 412–417 (2009).

58. L. V. Wang and S. Hu, Photoacoustic tomography: In vivo imaging from organelles to organs, Science **335**, 1458–1462 (2012).

59. A. Taruttis and V. Ntziachristos, Advances in real-time multispectral optoacoustic imaging and its applications, Nature Photonics **9**, 219 (2015).

60. L. V. Wang, J. J. Yao, A practical guide to photoacoustic tomography in the life sciences, Nat. Methods **13**, 627-638 (2016).

61. A. Bojahr, M. Herzog, D. Schick, I. Vrejoiu, and M. Bargheer, Calibrated real-time detection of nonlinearly propagating strain wave, Phys. Rev. B **86**, 144306 (2012).

62. C. Klieber, V. E. Gusev, T. Pezeril, and K. A. Nelson, Nonlinear acoustics at GHz frequencies in a viscoelastic fragile glass former, Phys. Rev. Lett. **114**, 065701 (2015).

63. T. Dehoux, K. Ishikawa, P. H Otsuka, M. Tomoda, O. Matsuda, M. Fujiwara, S Takeuchi, I. A Veres, V. E Gusev and O. B Wright, Optical tracking of picosecond coherent phonon pulse focusing inside a sub-micron object**,** Light: Science & Applications **5**, e16082 (2016).

64. V. Gusev, Laser hypersonics in fundamental and applied research, Acustica Acta Acustica **82**, S37 (1996).

65. V. Gusev, A. M. Lomonosov, P. Ruello, A. Ayouch, G. Vaudel, Depth-profiling of elastic and optical inhomogeneities in transparent materials by picosecond ultrasonic interferometry: Theory, J. Appl. Phys. 110, 124908 (2011).

66. J. W. Tucker and V. W. Rampton, *Microwave Ultrasonics in Solid State Physics* (North-Holland Publ. Co., Amsterdam, 1972).

67. Yu. A. Kravtsov and Yu. I. Orlov, *Geometrical Optics of Inhomogeneous Media* (Springer, Berlin, 1990).

68. L. M. Brekhovskikh and O. A. Godin, Acoustics of Layered Media I: Plane and Quasi-Plane Waves (Springer-Verlag, Berlin, 1990).

69. V. E. Gusev, Detection of nonlinear picosecond acoustic pulses by time-resolved Brillouin scattering, J. Appl. Phys. **116**, 064907 (2014).

70. P. Emery and A. Devos, Acoustic attenuation measurements in transparent materials in the hypersonic range by picosecond ultrasonics, Appl. Phys. Lett. 89, 191904 (2006).

71. A. Devos, G. Caruyer, Structure characterization based on wavelength effect in a photoacoustic system, 2005. PCT/FR 2006/001386.

72. J. K. Miller, J. Qi, Y. Xu, Y.-J. Cho, X. Liu, J. K. Furdyna, I. Perakis, T. V. Shahbazyan, and N. Tolk, Near-bandgap wavelength dependence of long-lived traveling coherent longitudinal acoustic phonons in GaSb-GaAs heterostructures, Phys. Rev. B **74**, 113313 (2006).

73. F. Hudert, A. Bartels, T. Dekorsy, and K. Köhler, Influence of doping profiles on coherent acoustic phonon detection and generation in semiconductors, J. Appl. Phys. **104**, 123509 (2008).

74. A. Devos, Colored ultrafast acoustics: From fundamentals to applications, Ultrasonics 56 (2015) 90–97

75. A. Devos, M. Foret, S. Ayrinhac, P. Emery, B. Ruffl, Hypersound damping in vitreous silica measured by picosecond acoustics, Phys. Rev. B **77** (2008).
100201.

76. S. Ayrinhac, M. Foret, A. Devos, B. Rufflé, E. Courtens, R. Vacher, Subterahertz hypersound attenuation in silica glass studied via picosecond acoustics, Phys. Rev. B **83** (2011) 014204.

77. A. Devos, J.-F. Robillard, R. Cote, P. Emery, High-laser-wavelength sensitivity of the picosecond ultrasonic response in transparent thin films, Phys. Rev. B **74**, 064114 (2006).

78. A. Nagakubo1, M. Arita, T. Yokoyama, S. Matsuda, M. Ueda, H. Ogi1, and M. Hirao, Acoustic properties of co-doped AlN thin films at low temperatures studied by picosecond ultrasonics, Jap. J. Appl. Phys. **54**, 07HD01 (2015).

79. C. Rossignol, N. Chigarev, M. Ducousso, B. Audoin, G. Forget, F. Guillemot, M.C. Durrieu, In Vitro picosecond ultrasonics in a single cell, Appl. Phys. Lett. **93**, 123901 (2008).

80. T. Dehoux, N. Tsapis, B. Audoin, Relaxation dynamics in single polymer microcapsules probed with laser-generated GHz acoustic waves, Soft Matter **8**, 2586 (2012).

81. T. Dehoux, M. Abi Ghanem, O. F. Zouani, M. Ducousso, N. Chigarev, C. Rossignol, N. Tsapis, M.-C. Durrieu, B. Audoin, Probing single-cell mechanics with picosecond ultrasonics, Ultrasonics **56** 160 (2015).

82. M. Khalizov, J. Pakarinen, L. He, H. B. Henderson, M. V. Manuel, A. T. Lelson, B. J. Jaques, D. P. Butt, D. H. Hurley, Subsurface imaging of grain microstructure using picosecond ultrasonics, Acta Materialia **112**, 209 (2016).

83. P. A. Elzinga, F. E. Lytle, Y. Jian, G. B. King, and N. M. Laurendeau, Pump/Probe spectroscopy by asynchronous optical sampling, Appl. Spectrosc. **41**, 2 (1987).

84. A. Bartels, R. Cerna, C. Kistner, A.Thoma, F. Hudert, C. Janke, T. Dekorsy, Ultrafast time-domain spectroscopy based on high-speed asynchronous optical sampling. Rev Sci Instrum. **78**, 035107 (2007).

85. K. H. Lin, C. T. Yu, Y. C. Wen, and C. K. Sun, Generation of picosecond acoustic pulses using a p-n junction with piezoelectric effects, Appl. Phys. Lett. **86**, 093110 (2005).

86. J. F. Ziegler, J. P. Biersack, and U. Littmark, *The Stopping and Ranges of Ions in Solids* (Pergamon, New York, 2000).

87. R. Cote, A. Devos, R. Cote, A. Devos, Refractive Index, Sound Velocity and Thickness of Thin Transparent Films from Multiple Angles Picosecond Ultrasonics. Rev. Sci. Instrum. **76**, 053906 (2005).

88. M. Tomoda, O. Matsuda, O. B. Wright, R. Li Voti, Tomographic reconstruction of picosecond acoustic strain propagation, Appl. Phys. Lett. **90**, 041114 (2007).

89. D. Ségur, Y. Guillet, and B. Audoin, Intrinsic geometric scattering probed by picosecond optoacoustics in a cylindrical cavity: Application to acoustic and optical characterizations of a single micron carbon fiber, Appl. Phys. Lett., 97, 031901 (2010).

90. H. N. Lin, H. J. Maris, L. B. Freund, K. Y. Lee, H. Luhn, and D. P. Kem, Study of vibrational modes of gold nanostructures by picosecond ultrasonics, J. Appl. Phys. 73, 37 (1993).

91. O. Matsuda, T. Pezeril, I. Chaban, K. Fujita, and V. Gusev, Time-domain Brillouin scattering assisted by diffraction gratings, Phys. Rev. B 97, 064301 (2018).

92. F. Decremps, L. Belliard, B. Perrin, M. Gauthier, Sound velocity and absorption measurements under high pressure using picosecond ultrasonics in a diamond anvil cell: application to the stability study of AlPdMn. Phys. Rev. Lett. **100**, 035502 (2008).

93. M. R. Armstrong, J. C. Crowhurst, E. J. Reed, J. M. Zaug. Ultrafast high strain rate acoustic wave measurements at high static pressure in a diamond anvil cell. Appl. Phys. Lett. **92**, 101930 (2008).

94. S. Hirsekorn, The scattering of ultrasonic waves by polycrystals. J. Acoust. Soc. Am. **72**, 1921 (1982).

95. C. M. Sayers, Ultrasonic velocities in anisotropic polycrystalline aggregates. J. Phys. D: Appl. Phys. **18**, 2157 (1982).

96. C.-S. Zha, R. J. Hemley, S. A. Gramsch, H.-K. Mao, W. A. Bassett, Optical study of $H_2O$ ice to 120 GPa: dielectric function, molecular polarizability, and equation of state. J. Chem. Phys. **126**, 074506 (2007).

97. M. Kuriakose, S. Raetz,Q. M. Hu, S. M. Nikitin, N. Chigarev, V. Tournat, A. Bulou, A. Lomonosov, P. Djemia, V. E. Gusev, and A. Zerr, Longitudinal sound velocities, elastic anisotropy, and phase transition of high-pressure cubic H2O ice to 82GPa, Phys. Rev. B **96**, 134122 (2017).

98. L. Shelton, F. Yang, W. Ford, H. Maris, Picosecond ultrasonic measurement of the velocity of phonons in water, Phys. Status Solidi B **242**, 1379 (2005).

99. A.A. Maznev, K.J. Manke, C. Klieber, K.A. Nelson, S.H. Baek, C.B. Eom, Coherent Brillouin spectroscopy in a strongly scattering liquid by picosecond ultrasonics, Opt. Lett. **36** 2925 (2011).

96. C. Rossignol, N. Chigarev, M. Ducousso, B. Audoin, G. Forget, F. Guillemot, M.C. Durrieu, In Vitro picosecond ultrasonics in a single cell, Appl. Phys. Lett. **93**, 123901 (2008).

100. O. F. Zouani, T; Dehoux, M.-C. Durrieu and B Audoin,, Universality of the network-dynamics of the cell nucleus at high frequencies, Soft Matter, **10**, 8737 (2014).

101. T. Dehoux, B. Audoin, Non-invasive optoacoustic probing of the density and stiffness of single biological cells, J. Appl. Phys. **112**, 124702 (2012).

102. B. Audoin, C. Rossignol, N. Chigarev, M. Ducousso, G. Forget, F. Guillemot, M.C. Durrieu, Picosecond acoustics in vegetal cells: non-invasive in vitro measurements at a sub-cell scale, Ultrasonics **50**, 202 (2010).

103. A. Gadalla, T. Dehoux, and B. Audoin, Transverse mechanical properties of cell walls of single living plant cells probed by laser-generated acoustic waves, Planta 239, 1129 (2014).

104. R. J. Smith, F. Perez-Cota, L. Marques, X. Chen, A. Arca, K. Webb, J. Aylott, M. G. Somekh, and M. Clark, Optically excited nanoscale ultrasonic transducers, J. Acoust. Soc. Am. **137**, 219 (2015).

105. A. V. Akimov, E. S. K. Young, J. S. Sharp, V. Gusev, A. J. Kent, Coherent hypersonic closed-pipe organ like modes in supported polymer films, Appl. Phys. Lett. **99**, 021912 (2011).

106. C. Klieber,T. Hecksher, T. Pezeril, D. H. Torchinsky,J. C. Dyre, and K. A. Nelson, Mechanical spectra of glass-forming liquids. II. Gigahertz-frequency longitudinal and shear acoustic dynamics in glycerol and DC704 studied by time-domain Brillouin scattering, J. Chem. Phys. **138**, 12A544 (2013).

107. F. Yang, T. J. Grimseley, S. Che, G. A. Antonelli, H. J. Maris, and A. V. Nurmikko, Picosecond ultrasonic experiments with water and its application to the measurement of nanostructures, J. Appl. Phys. **107**, 103537 (2010).

108. K. Yu, T. Devkota, G. Beane, G. P. Wang, and G. V. Hartland, Brillouin oscillations from single Au nanoplate opto-acoustic transducers, ACS Nano **11**, 8064–8071 (2017).

109. S. Che, P.R. Guduru, A.V. Nurmikko, H.J. Maris, A scanning acoustic microscope based on picosecond ultrasonics, Ultrasonics **56**, 153 (2015).

110. K-H. Lin, C-M. Lai, C-C. Pan, J-I. Chyi, J-W. Shi, S-Z. Sun, C-F. Chang, C.-K. Sun, Spatial manipulation of nanoacoustic waves with nanoscale spot sizes. Nat Nanotechnol **2**, 704 (2007).

111. A. Vertikov, M. Kuball, A. V. Nurmikko, H. J. Maris, Time-resolved pump-probe experiments with subwavelength lateral resolution, Appl. Phys. Lett. **69**, 2465 (1996).

112. T. Bienville, L. Belliard, P. Siry, B. Perrin, Photothermal experiments in the time and frequency domains using an optical near field microscope, Superlatttices **35**, 363 (2004).

113. O. V. Rudenko and S. I. Soluyan, *Theoretical Foundations of Nonlinear Acoustics* (Consultant Bureau, New York, 1977).

114. O. L. Muskens and J. I. Dijkhuis, High amplitude, ultrashort, longitudinal strain solitons in sapphire, Phys. Rev. Lett. **89**, 285504 (2002).

115. R. J. Smith, R. A. Light, S. D. Sharples, N. S. Johnston, M. C. Pitter, and M. G. Somekh, Multichannel, time-resolved picosecond laser ultrasound imaging and spectroscopy with custom complementary metaloxide- semiconductor detector, Rev. Sci. Instrum. **81**, 024901 (2010).

116. T. Pezeril, C. Klieber, V. Shalagatskyi, G. Vaudel, V. Temnov, O. G. Schmidt, and D. Makarov, Femtosecond imaging of nonlinear acoustics in gold, Optics Express **22**, 4590-4598 (2014).

117. C. Brüggemann, A. V. Akimov, A. V. Scherbakov, M. Bombeck, C. Schneider, S. Höfling, A. Forchel, D. R. Yakovlev, and M. Bayer, Laser mode feeding by shaking quantum dots in a planar microcavity, Nature Photonics **6**, 30 (2012).

118. T. Czernik, C. Schneider, M. Kamp, S. HOfling, B. A. Glavin, D. R. Yakovlev, A. V. Akimov, M. Bayer, Acousto-optical nanoscopy of buried photonic nanostructures, Optica **4**, 2334 (2017).

119. B. Perrin, B. Bonello, J. C. Jeannet, and E. Romatet, Prog. Nat. Sci. **S6**, 444 (1996).

120. H. Hurley and O. B. Wright, Detection of ultrafast phenomena by modified Sagnac interferometer, Opt. Lett. **24**, 1305 (1999).

121. A. V. Scherbakov, M. Bombeck, J. V. Jäger, A. S. Salasyuk, T. L. Linnik, V. E. Gusev, D. R. Yakovlev, A. V. Akimov, and M. Bayer, Picosecond opto-acoustic interferometry and polarimetry in high-index GaAs, Optics Express **21**, 16473 (2013).

122. O. Matsuda, O. B. Wright, D. H. Hurley, V. Gusev, and K. Shimizu, Coherent shear phonon generation and detection with picosecond laser acoustics, Phys. Rev. B **77**, 224110 (2008).

123. D. Mounier, E. Morozov, P. Ruello, M. Edely, P. Babilotte, C. Mechri, J.-M. Breteau, and V. Gusev, Application of transient grating femtosecond polarimetry/ellipsometry technique in picosecond laser ultrasonics, J. Phys.: Conf. Ser. **92**, 012179 (2007).

124. D. Mounier, E. Morozov, P. Ruello, J.-M. Breteau, P. Picart, and V. Gusev, Detection of shear picosecond acoustic pulses by transient femtosecond polarimetry, Eur. Phys. J. Spec. Top. **153**, 243 (2008).

125. D. Mounier, P. Picart, P. Babilotte, P. Ruello, J.-M. Breteau, T. Pezeril, G. Vaudel, M. Kouyate, and V. Gusev, Jones matrix formalism for the theory of picosecond shear acoustic pulse detection, Opt. Express **18**, 6767 (2010).

126. M. Lejman, G. Vaudel, I. C. Infante, I. Chaban, T. Pezeril, M. Edely, G. F. Nataf, M. Guennou, J. Kreisel, V. E. Gusev, B. Dkhil, and P. Ruello, Ultrafast acousto-optic mode conversion in optically birefringen ferroelectrics, Nature Communications **7**, 12345 (2016).

127. A. Baydin, H. Krzyzanowska, R. Gatamov, J. Garnett, N. Tolk, The photoelastic coefficient $P_{12}$ of H+ implanted GaAs as a function of defect density, Scientific Reports 7, 15150 (2017).

128. T. Bienville and B. Perrin, Generation and detection of quasi transverse waves in an anisotropic crystal by picosecond ultrasonics, in "Proceedings of the World congress on ultrasonics : WCU 2003, Paris, France, 7-10 September 2003," (Société française d'acoustique, Paris, 2003), pp. 813–816.

129. T. Pezeril, N. Chigarev, P. Ruello, S. Gougeon, D. Mounier, J.-M. Breteau, P. Picart, and V. Gusev, Laser acoustics with picosecond collimated shear strain beams in single crystals and polycrystalline materials, Phys. Rev. B **73**, 132301 (2006).

130. R. N. Kini, A. J. Kent, N. M. Stanton, and M. Henini, Generation and detection of terahertz coherent transverse-polarized acoustic phonons by ultrafast optical excitation of gaas/alas superlattices, Appl. Phys. Lett. **88**, 134112 (2006).

131. T. Pezeril, C. Klieber, S. Andrieu, and K. A. Nelson, Optical generation of gigahertz-frequency shear acoustic waves in liquid glycerol, Phys. Rev. Lett. **102**, 107402 (2009).

132. M. Lejman, G. Vaudel, I. C. Infante, V. E. Gusev, B. Dkhil, and P. Ruello, Giant ultrafast photo-induced shear strain in ferroelectric $BiFeO_3$, Nat. Commun. **5**, 4301 (2014).

133. O. Matsuda and O. B. Wright, Theory of detection of shear strain pulses with laser picosecond acoustics, Anal. Sci. **17**, S216 (2001).

134. A. van der Drift, Evolutionary selection, a principle governing growth orientation in vapour-deposited layers, Philips Res. Rep. **22**, 267 (1967).

135. P. Belleville, C. Bonnin, J. J. Priotton, Room-temperature mirror preparation using sol-gel chemistry and laminar-flow coating technique. J. Sol-Gel. Sci. Technol. **19**, 223 (2000).

136. J. Dai, P. Mukundhan, C. Kim, and H. J. Maris, Analysis of a picosecond ultrasonic method for measurement of stress in a substrate, J. Appl. Phys. **119**, 105705 (2016).

137. S. Kashiwada, 0. Matsuda, J. J. Baumberg, R. Li Voti and O. B. Wright, In situ monitoring of the growth of ice films by laser picosecond acoustics, J. Appl. Phys. **100**, 073506 (2006).

138. M. Kuriakose, N. Chigarev, S. Raetz, A. Bulou, V. Tournat, A. Zerr and V. E. Gusev, In situ imaging of the dynamics of photo-induced structural phase transition at high pressures by picosecond acoustic interferometry, New J. Phys. **19**, 053206 (2017).